\documentclass[12pt,a4paper]{article}
\usepackage{latexsym,graphicx,psfrag,amsmath}
\def\fn#1{\footnotesize\hspace*{-0.5em} #1}%
\def\gn#1{\hspace*{3em} \pageref{#1}}%

\newcommand{\im}{\mbox{Im}}
\newcommand{\arctanh}{\mbox{arctanh}}

\newcommand{\be}{\begin{equation}}
\newcommand{\bdm}{\begin{displaymath}}
\newcommand{\edm}{\end{displaymath}}
\newcommand{\bea}{\begin{eqnarray}}
\newcommand{\eea}{\end{eqnarray}}

\newcommand{\fs}{\; .}
\newcommand{\co}{\; ,}
\newcommand{\al}{\!&\!}
\newcommand{\eff}{{e\hspace{-0.1em}f\hspace{-0.18em}f}}
\newcommand{\ghb}{\gamma_1}
\newcommand{\gth}{\gamma_2}
\newcommand{\J}{H}
\newcommand{\JS}{I}
\newcommand{\JR}{R}

\newcommand{\Jbar}{\bar{H}}
\newcommand{\JSbar}{\bar{I}}
\newcommand{\JRbar}{\bar{R}}
\newcommand{\cS}{i\rule{0.05em}{0em}}

\newcommand{\Jdiv}{\nu}
\newcommand{\lambdabar}{\bar{\lambda}}
\newcommand{\Mbare}{m}
\newcommand{\MN}{m_N}
\newcommand{\Mp}{M}
\newcommand{\gAbare}{g_A}
\newcommand{\sth}{s_+}
\newcommand{\wave}{\raisebox{0.15em}{\fbox{\rule[0.15em]{0em}{0em}\,}}\,}

\newcommand{\lambdaN}{\lambdabar}

\newcommand{\Pslash}{P\hspace{-0.65em}\slash\,}
\newcommand{\Dpi}{\Delta_\pi}
\newcommand{\DN}{\Delta_N}
\newcommand{\sigmaN}{\sigma}
\newcommand{\bphi}{\mbox{\boldmath$\pi$}}
\newcommand{\ch}{\protect\mbox{\hspace{0.02em}\protect\raisebox{0.23em}
{$\chi$}}}
\newcommand{\bHBCHPT}{\boldmath HB\protect\raisebox{0.2em}{$\chi$}PT}
\newcommand{\GSS}{\mbox{\tiny\it GSS}}
\begin{document}

\begin{flushright}{\large BUTP--99/1\rule{1em}{0em}}\end{flushright}

\vspace{6em}
\begin{center}
{\LARGE\bf Baryon Chiral Perturbation Theory}

\vspace{0.4em}
{\LARGE\bf in Manifestly Lorentz Invariant
  Form}

\vspace{4em} 
 T.~Becher and H.~Leutwyler\\
Institute for Theoretical Physics, University of Bern\\
Sidlerstr.~5, CH-3012 Bern, Switzerland
\end{center}

\vspace{2em}
\begin{abstract} {We show that in the presence of massive particles such as
nucleons, the standard low energy 
expansion in powers of meson momenta and light quark
masses in general only converges in part of the low energy region.
The expansion of the scalar form factor $\sigma(t)$, for instance, 
breaks down in the 
vicinity of $t=4M_\pi^2$. In the language of heavy baryon 
chiral perturbation 
theory, the proper behaviour in the threshold region only results if the
multiple internal line insertions generated by relativistic kinematics are
summed up to all orders. We propose a method that yields a coherent
representation throughout the low energy region while keeping Lorentz and
chiral invariance explicit at all stages. The method is illustrated 
with a calculation of the nucleon mass and of the scalar form factor to order 
$p^4$.} 
\end{abstract}

\begin{center}
\vspace{10em}
\rule{28em}{0.02em}\\
{\footnotesize Work supported in part by Schweizerischer Nationalfonds}  
\end{center}
\thispagestyle{empty}
\newpage
\thispagestyle{empty}

\section*{\hspace{8em}Contents}

\begin{tabular}{rlr}
1\al Introduction & \gn{intro}\\
2\al Effective Lagrangian & \gn{eL}\\
3\al Scalar form factor & \gn{scalar}\\
4\al Low energy expansion near threshold & \gn{threshold}\\
\vspace*{-0.2em}
5\al Infrared singularities in the self energy & \gn{selfenergy}\\
\vspace*{-0.2em}
&\fn{Singular and regular parts} & \gn{se1}\\
\vspace*{-0.2em}
&\fn{Properties of the decomposition} & \gn{se2}\\
\vspace*{-0.2em}
&\fn{Representation in terms of modified propagators}&\gn{se3}\\
&\fn{Dispersive representation}&\gn{se4}\\\vspace*{-0.2em}
6\al Generalization&\gn{generalization}\\\vspace*{-0.2em}
&\fn{Singular and regular parts}&\gn{gen1}\\
&\fn{Uniqueness of the decomposition}&\gn{gen2}\\\vspace*{-0.2em}
7\al Comparison with HB\ch PT&\gn{comparison}\\\vspace*{-0.2em}
&\fn{Lorentz invariance}&\gn{hb1}\\\vspace*{-0.2em}
&\fn{Chiral symmetry}&\gn{hb2}\\
&\fn{Infrared part as an alternative regularization}&\gn{hb3}\\
\vspace*{-0.2em}
8\al Renormalization&\gn{renormalization}\\\vspace*{-0.2em}
&\fn{Self energy} &\gn{ren1}\\\vspace*{-0.2em}
&\fn{Renormalization scale}&\gn{ren2}\\
&\fn{Other one loop graphs}&\gn{ren3}\\\vspace*{-0.2em}
9\al Convergence of the chiral expansion, explicit representations&
\gn{convergence}\\
\vspace*{-0.2em}
&\fn{Explicit representation of the self energy}&\gn{con1}\\
\vspace*{-0.2em}
&\fn{Chiral expansion of the self energy}&\gn{con2}\\
&\fn{Other one loop integrals}&\gn{con3}\\
10\al Chiral expansion of the nucleon mass&\gn{nucleon mass}\\
11\al Wave function renormalization &\gn{wave function}\\
\vspace*{-0.2em}
12\al Scalar form factor to order $p^4$&\gn{formfactor}\\
\vspace*{-0.2em}
&\fn{Evaluation of the graphs}&\gn{sff1}\\\vspace*{-0.2em}
&\fn{Result}&\gn{sff2}\\\vspace*{-0.2em}
&\fn{Unitarity}&\gn{sff3}\\\vspace*{-0.2em}
&\fn{Value at the Cheng-Dashen point}&\gn{sff4}\\
&\fn{The $\sigma$-term}&\gn{sff5}\\\vspace*{-0.2em}
13\al Discussion &\gn{discussion}\\\vspace*{-0.2em}
&\fn{Reordering of the perturbation series}&\gn{disc1}\\
\vspace*{-0.2em}
&\fn{The role of the $\Delta(1232)$}&\gn{disc2}\\\vspace*{-0.2em}
&\fn{Comparison with the static model}&\gn{disc3}\\
&\fn{Momentum space cutoff}&\gn{disc4}\\
14\al Conclusion& \gn{conclusion}\\
A\al Low energy representation for the triangle graph & \gn{A}\\
B\al Infrared parts of some loop integrals & \gn{Loop integrals}\\
C\al $\pi N$ scattering in tree approximation &
\gn{scattering}\\
D\al Contributions generated by the $\Delta$ &\gn{delta}
\end{tabular}

\newpage
\setcounter{page}{1}

\section{Introduction}\label{intro}
The effective low energy theory of the strong
interaction is based on a simultaneous expansion of the Green functions of
QCD in powers of the external momenta and of the quark masses (``chiral
expansion''). In the vacuum
sector, where the only low energy singularities are those generated by the 
Goldstone bosons, dimensional regularization yields homogeneous functions of
the momenta and Goldstone masses, so that each graph has an unambiguous order
in the chiral expansion. In the sector with baryon number 1,
however, the 
low energy structure is more complicated. The corresponding effective theory 
can be formulated in manifestly Lorentz invariant form \cite{gss},
but it is not a trivial matter to keep track of the
chiral order of graphs containing loops within that framework: The chiral 
expansion of the loop graphs in general starts at the same order as the
corresponding tree graphs, so that 
the renormalization of the divergences requires a tuning also of those 
effective coupling constants that occur at lower order -- in particular, the
nucleon mass requires renormalization at every order of the series. 

Most of the recent calculations avoid this complication with a 
method referred to as heavy baryon chiral perturbation theory (HB\ch PT)
\cite{man,mei,EckerReview,mei2}. The 
starting point of that method is the same Lorentz invariant effective 
Lagrangian that occurs in the relativistic approach. The loop graphs, however,
are evaluated differently: The Dirac-spinor
that describes the nucleon degrees of freedom is reduced to a two-component 
field and the baryon kinematics is expanded around the nonrelativistic
limit. 
At the end of the calculation, the amplitude may then be recast into Lorentz
invariant form. This method keeps track of the chiral power counting at every 
step of the calculation, at the price of manifest Lorentz invariance. 

Moj\v zi\v s \cite{moj} and Fettes et al.\ \cite{Fettes} have evaluated 
the $\pi N$
scattering amplitude to order $p^3$ within that framework. The explicit result
is remarkably simple, the one loop graphs being expressible in terms of
elementary functions. On general grounds, the outcome of this calculation must
be the same as what is obtained if the representation of the scattering
amplitude given in \cite{gss} is expanded around the nonrelativistic limit --
that representation also holds to order $p^3$. Indeed, Ellis and Tang
\cite{Tang} have shown that this is the case.

Quite apart from the fact that the nonrelativistic expansion turns the
effective Lagrangian into a rather voluminous object and that care is required
to properly analyze the corresponding loop graphs\footnote{Wave function
  renormalization, for instance, is not a trivial matter \cite{wave function
    renormalization,Ecker,Steininger,Kambor Mojzis}.}, HB\ch PT suffers from a
deficiency: The corresponding perturbation series fails to converge in part of
the low energy region. The problem is generated by a set of higher order
graphs involving insertions in nucleon lines. A similar phenomenon also
appears in the effective field theory of the $NN$-interaction
\cite{NN}. It arises
from the nonrelativistic expansion and does not occur in the relativistic
formulation of the effective theory.
  
The purpose of the present paper is to present a method that exploits the
advantages of the two techniques and avoids their disadvantages. In sections 3
and 4, we demonstrate the need to sum up certain graphs of HB\ch PT, using the
example of the scalar nucleon form factor. Next, we show that the infrared
singularities of the various one loop graphs occurring in the chiral
perturbation series can be extracted in a relativistically invariant fashion
(sections 5 and 6). The method we are using here follows the approach of Tang
and Ellis \cite{Tang}. There is a slight difference, insofar as we do not rely
on the chiral expansion of the loop integrals -- that expansion does not
always converge. A comparison with HB\ch PT is given in section 7, where we
also show that our procedure may be viewed as a novel method of
regularization, which we call ``infrared regularization''. The method
represents a variant of dimensional regularization. While that scheme permits
a straightforward counting of the powers of momentum at high energies,
infrared regularization preserves the low energy power counting rules that
underly chiral perturbation theory.  Renormalization is discussed in section
8. We then analyze the convergence of the chiral expansion of the one loop
integrals and show that the expansion coefficients relevant for $\pi N$
scattering can be expressed in terms of elementary functions (section 9).
Whenever that expansion converges, the result agrees with the one obtained
within HB\ch PT. We also construct an explicit low energy representation of
the triangle graph, for which the nonrelativistic expansion fails. The method
is illustrated with an evaluation of the chiral perturbation series for the
nucleon mass, the wave function renormalization constant and the scalar form
factor. The physics of the result obtained for the form factor and for the
$\sigma$-term is discussed in sections 12 and 13. We demonstrate that a
significant part of those infrared singularities that are proportional to
$M_\pi^3$ is common to the form factor and to the $\pi N$ scattering amplitude
and thus drops out when considering the predictions of the theory. The
observation leads to a specific reordering of the chiral perturbation series
that reduces the matrix elements of the perturbation quite substantially.  The
effects generated by the $\Delta(1232)$ are discussed in some detail (appendix
\ref{delta}) and we also compare our framework with the ancestor of the
effective theory described here: the static model. Section 14 contains our
conclusions.

\section{Effective Lagrangian}\label{eL}
The variables of the effective theory are the meson field
$U(x)\in\mbox{SU(2)}$ and the Dirac spinor $\psi(x)$ describing the degrees of
freedom of the nucleon. The effective Lagrangian contains two pieces,
\bea {\cal L}_\eff={\cal L}_\pi +{\cal L}_N\fs\nonumber\eea
The first part is the well-known meson Lagrangian,
which only involves the field $U(x)$ and an even number of derivatives
thereof. For the second part, which
is bilinear in $\bar{\psi}(x)$ and $\psi(x)$, the derivative expansion
contains odd as well as even terms:
\bea{\cal L}_N
={\cal L}_N^{(1)} +{\cal L}_N^{(2)}+{\cal L}_N^{(3)}+\ldots\nonumber\eea
The explicit expressions involve
the quantities $u$, $u_\mu$, $\Gamma_\mu$ and $\chi_{\pm}$.
In the absence of external vector and axial fields, these are given by
\bea  u^2 \al=\al U\co\hspace{1.5em}
u_\mu =i u^\dag\partial_\mu U u^\dag\co\hspace{1.5em}
\Gamma_\mu = \mbox{$\frac{1}{2}$}[u^\dag,\partial_\mu u]\co\hspace{1.5em}
\chi_\pm = u^\dag \chi u^\dag \pm u \chi^\dag u 
\nonumber\fs\eea
At leading order, the effective Lagrangian is fully determined by the
nucleon mass and by the matrix element of the axial 
charge ($\Mbare$ and $\gAbare$ denote the corresponding leading order values
and $D_\mu\equiv\partial_\mu+\Gamma_\mu$):
\be {\cal L}^{(1)}_{N}=\bar \psi\left( iD \!\!\!\!/-\Mbare\right)\psi +
\mbox{$\frac{1}{2}$}\, \gAbare\,
\bar{\psi}\, u \hspace{-0.5em}\slash\,\gamma_5 \psi\co\nonumber\end{equation}
We disregard isospin breaking effects.
The Lagrangian of order $p^2$ then 
contains four independent coupling constants\footnote{We use the conventions
  of ref.~\cite{mei}. In this notation,
the coupling constants of ref.~\cite{gss}
are given by: $\Mbare\, c_1^{\GSS}\! =\! F^2 c_1$, $\Mbare\, c_2^{\GSS}\!=\! - 
F^2 c_4$, 
$\Mbare\,  c_3^{\GSS}\!=\! - 2 F^2 c_3$, $\Mbare\, (c_4^{\GSS}-2\,\Mbare\, 
c_5^{\GSS})\!=\!- F^2 c_2$ (to order $p^2$, the terms 
$c_4^{\GSS}$ and $c_5^{\GSS}$  enter the observables only in
  this combination), while those of 
ref.~\cite{EckerRenormierung} read:  
$16\, a_1\!=\! 8\, \Mbare
\,  c_3+\gAbare^2$, $8\, a_2\! =\! 4\,\Mbare\, c_2-\gAbare^2$, $a_3\!=\! 
\Mbare\, c_1$,  $4\,a_5\! =\! 4\,\Mbare\, c_4+1-\gAbare^2$. 
In the numerical analysis, we work with $F_\pi=92.4\,\mbox{MeV}$, $g_A=1.267$,
$M_\pi=M_{\pi^+}$.}
\bea  \label{L2}
{\cal L}^{(2)}_{N}\al=\al c_1\, \langle\chi_+\rangle\,\bar \psi\, \psi
  -\frac{c_2}{4\Mbare^2}\,\langle u_\mu u_\nu \rangle\,
(\bar{\psi}\,  D^\mu\! D^{\nu}\!\psi + \mbox{h.c.})\,\\\al\al
+\frac{c_3}{2}\,\langle u_\mu u^\mu\rangle\,\bar{\psi}\,\psi
 -\frac{c_4}{4}\,\bar{\psi}\,\gamma^\mu\,\gamma^\nu\, [u_\mu,u_\nu]\,\psi
\fs\nonumber\eea
Below, we apply the machinery to the scalar form factor. 
This quantity does not receive any contribution from ${\cal  L}^{(3)}_{N}$, 
but two of the terms in ${\cal L}^{(4)}_N$,
\bea\label{Lag4} 
{\cal L}_N^{(4)}=-\frac{e_1}{16}\langle\chi_+\rangle^2 \bar{\psi}\,\psi+
\frac{e_2}{4}\,\langle \chi_+ \rangle\, \wave(\bar{\psi}\,\psi)+\ldots
\;\co\eea
do generate contributions proportional to $e_1 M_\pi^4$ and $e_2 M_\pi^2\,t$, 
respectively.
\section{Scalar form factor}\label{scalar}
We first wish to show that, in the sector with baryon number 1, 
the standard chiral expansion in powers
of meson momenta and quark masses converges only in part of the low energy
region. For definiteness, we consider the scalar form factor of the nucleon in
the isospin limit ($m_u=m_d=\hat{m}$),
\begin{equation}
  \langle N(P',s')|\,\hat{m}\, (\bar u u+ \bar d d)\,| N(P,s)\rangle=\bar u'
  u\; \sigma(t),\hspace{2em} t=(P'-P)^2\fs \nonumber
\end{equation}
The first two terms occurring in the low energy expansion of this form factor 
were worked out long ago, on the basis of a one loop calculation within the
Lorentz invariant formulation of the effective theory \cite{gss}. In that
expansion,  $t$, $\hat{m}$ and $M_\pi^2$ 
are treated as small quantities of $O(p^2)$, 
while the nucleon mass represents
 a term of $O(p^0)$. In view of the quark mass factor occurring in the
 definition of $\sigma(t)$, the low energy expansion starts at order
 $p^2$, with a momentum independent term generated by 
${\cal L}_N^{(2)}$:
\begin{equation}\label{scalar form factor}
\sigma(t)= -4 c_1\,M_\pi^2 +
\frac{3\, g_A^2 M_\pi^2 \MN }{4F_\pi^2}
\left\{(t-2M_\pi^2)\,\gamma(t)-\frac{M_\pi}{8\pi \MN}\right\}+O(p^4)
\end{equation}
The constant $c_1$ occurring here is a renormalized version of the bare
coupling constant in eq.~(\ref{L2}). Since the renormalization depends on the
framework used, we do not discuss it at this preliminary stage.  The
contribution of order $p^3$ is generated by the triangle graph shown in fig.~1
and is fully determined by $F_\pi$ and $g_A$.
\begin{figure}[htbp]
  \begin{center}
    \leavevmode
    \includegraphics[width=1.9in]{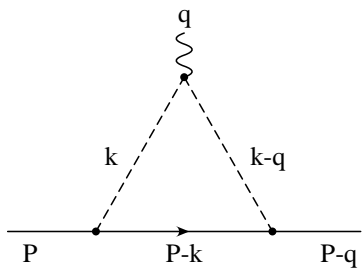}
    \caption{Triangle graph. The solid, dashed and wiggly lines
represent nucleons, pions and an external scalar source, respectively. }
    \label{fig:triangle}
  \end{center}
\vspace*{-1.5em}
\end{figure}
The term involves the convergent scalar loop integral
\bea\label{gammaloop}
\gamma(t)=\frac{1}{i}\!\int\!\!\frac{d^4 k}{(2\pi)^4}\frac{1}
{(\Mp^2\!-k^2\!-\!i\epsilon)\,
(\Mp^2\!-\!(k\!-\!q)^2\!-\!i\epsilon)\,
(\Mbare^2\!-\!(P\!-\!k)^2\!-\!i\epsilon)}\eea
Here and in the following,
we identify the masses occurring in the loop
integrals with their leading order values, $M_\pi\rightarrow M$,
$m_N\rightarrow m$. 

The function $\gamma(t)$ represents a
quantity of $O(1/p)$. Since the external nucleon lines are on the mass
shell, it exclusively depends on $t=q^2,\, \Mp$ and $\Mbare$. The function
is analytic in $t$ except 
for a cut along the positive real axis, starting at
$t=4\,\Mp^2$. The triangle graph also shows up in the analysis of the
$\pi N$ scattering amplitude to one loop, so that
the function $\gamma(t)$ is relevant also for that case.

The imaginary part of $\gamma(t)$ can be expressed in terms of elementary
functions \cite{gss}:  
\begin{equation}\label{imgamma}
\im \gamma(t)= 
\frac{\theta(t-4 \Mp^2)}{8\pi\sqrt{\,t\,(4\Mbare^2-t)}}\;\arctan
\frac{\sqrt{(t-4\Mp^2)(4\Mbare^2-t)}}{t-2\Mp^2}\fs
\end{equation}
Dropping corrections of order $t/\Mbare^2=O(p^2)$, this 
expression simplifies to
\begin{equation}\label{im gamma} \im \gamma (t)
 =\frac{\theta(t-4 \Mp^2)}{16\pi \Mbare\sqrt{t}}\left\{\arctan
\frac{2 \Mbare\sqrt{\,t-4\Mp^2}}{t-2\Mp^2}+O(p^2)\right\}\fs \end{equation}
The problem addressed above shows up in this formula:
The quantity
\bea
x=\frac{2 \Mbare\sqrt{\,t-4\Mp^2}}{t-2\Mp^2}\nonumber
\eea represents a term of $O(1/p)$. The standard
chiral expansion of $\mbox{Im}\gamma(t)$ thus corresponds to the series
$\arctan x = \frac{1}{2}\,\pi -1/x + \frac{1}{3}\, (1/x)^3 +\ldots $ 
This series, however, only
converges for $|x|>1$. In the vicinity of $t= 4 \Mp^2$, the
condition is not met, so that the chiral expansion diverges. 
The problem arises because the quantity $x$ takes small values there,
while the low energy expansion treats it as a large term of $O(1/p)$. 
In the region $|x|<1$, we may instead use the convergent
series 
$\arctan x= x - \frac{1}{3}\, x^3 +\ldots \;$, but this amounts to an
expansion in inverse powers of $p$. 

The rapid variation of the form factor near $t=4M^2$ is related to the fact
that the function $\arctan z$ exhibits branch points at $z=\pm\, i$. The
analytic continuation of $\gamma(t)$ to the second sheet therefore contains a
branch point just below threshold: \bea
\frac{(t-4\Mp^2)(4\Mbare^2-t)}{(t-2\Mp^2)^2}=-1 \hspace{2em}\rightarrow
\hspace{2em} t=4\Mp^2-\frac{\Mp^4}{\Mbare^2} \fs \nonumber\eea This implies
that, in the threshold region, the form factor does not admit an expansion in
powers of meson momenta and quark masses.
As was shown in ref.~\cite{mei}, the heavy
baryon perturbation series to $O(p^3)$ 
coincides with the chiral 
expansion of the relativistic result \cite{gss} and it was noted in 
ref.~\cite{mei2} that this representation does not make sense
near $t= 4 \Mp^2$. The corresponding
imaginary part amounts to the approximation 
$\arctan x\rightarrow\frac{1}{2}\,\pi$, so that the singularity
structure on the second sheet is discarded. Within HB\ch PT,
an infinite series of internal line insertions must be summed up to properly
describe the behaviour of the form factor near threshold.  The relativistic
formula (\ref{scalar form factor}), on the other hand, does apply in the
entire low energy region, because it involves the full function $\gamma(t)$
rather than the first one or two terms in the chiral expansion thereof.

\section{Low energy expansion near threshold}\label{threshold}
We conclude that in the threshold region, the low energy structure cannot be
analyzed in terms of the standard chiral expansion. The first two terms of
this expansion, i.e. the first two terms of the heavy
baryon perturbation series,
\bea \im \ghb (t)
 =\frac{\theta(t-4 \Mp^2)}{32\pi \Mbare \Mp }\left\{
\frac{\pi \Mp}{\sqrt{t}} -
   \frac{\alpha\,(t-2\,\Mp^2)}{\sqrt{\,t\,(t-4 \Mp^2)}}\right\}\co 
\eea
provide a decent representation only if $t$ is not close to threshold. 

To resolve the structure
in the threshold region, we need to consider an expansion that does not 
treat the quantity $x$ as large. This can be done by replacing the variable
$t$ with the dimensionless parameter
\bea
\xi=\frac{\sqrt{\,t-4 \Mp^2}}{\alpha\, \Mp}\co\hspace{2em}
 \alpha =\frac{\Mp}{\Mbare}\co\nonumber\eea
and expanding the amplitude at fixed $\xi$, so that the momentum transfer
\begin{equation} t=(4 + \alpha^2 \xi^2)\,M^2 \nonumber\end{equation}
stays close to threshold. 

For the imaginary part, the expansion at fixed $\xi$ takes the form
\begin{equation} \im \gamma (t)
 =\frac{\theta(t-4 \Mp^2)}{32\pi \Mbare \Mp}\left\{\arctan\xi
+O(p^2)\right\}\fs \nonumber\end{equation}
At large values of $\xi$, this representation smoothly joins the one provided
by the heavy baryon expansion, where $t/\Mp^2$ is kept fixed.
To amalgamate the two, we note that, in the threshold region, the quantity
$\im \ghb(t)$ reduces to the first two terms of the 
series $\arctan \xi\equiv\frac{1}{2}\,\pi -\arctan 1/\xi=
 \frac{1}{2}\,\pi - 1/\xi +\ldots$ 
Hence the difference
between $\im \gamma(t)$ and the first two terms of the heavy baryon
perturbation series is approximately given by 
\begin{equation}\label{imth} \im \gth (t)
 = \frac{\theta(t-4 \Mp^2)}{32\pi \Mbare \Mp }
\left\{   \frac{\alpha\, \Mp}{\sqrt{\,t-4 \Mp^2}}
 -\arctan\frac{\alpha\, \Mp}{\sqrt{\,t-4 \Mp^2}}\right\}\fs \end{equation}
Outside the threshold region, $\im \gth(t)$ 
is negligibly small -- in the chiral counting of 
powers, 
it represents a term
of order $p^2$. The representation
\begin{equation}\label{repim}\im \gamma(t)=\im \ghb(t) + 
\im \gth(t)+O(p)
\end{equation}
holds irrespective of whether $t/\Mp^2$ or $\xi$ is held fixed. 
Indeed, one may verify that on
the entire interval $4 \Mp^2 \leq t \leq 20\, \Mp^2$, 
the formula differs from
the expression in eq.~(\ref{im gamma}) by less than 1\,\%. 

Finally, we turn to the function $\gamma(t)$ itself. As mentioned above, the
loop integral cannot be expressed in terms of elementary functions. We may,
however, give an explicit representation that holds to first nonleading 
order of
the low energy expansion, throughout the low energy region. The calculation is
described in appendix A. It leads to a representation of the form
\bea\label{g12}\gamma(t)=\ghb(t)+\gth(t)+O(p)\fs\eea
The first term corresponds to
the result obtained in HB\ch PT \cite{mei}. This piece explodes in the 
vicinity of the threshold, like its imaginary part. The explicit expression for
the remainder reads
\begin{equation}
\gth(t)=\frac{1}{32\pi \Mbare \Mp}\left\{\frac{\alpha\,\Mp}
{\sqrt{4 \Mp^2-t}}-\mbox{ln}\left\{1+\frac{\alpha\,\Mp}
{\sqrt{4 \Mp^2-t}}\right\}\right\}\fs \end{equation}
In the language of HB\ch PT, this term is generated by multiple
internal line insertions. It takes significant values only in the immediate
vicinity of threshold, where it cures the deficiencies of $\ghb(t)$. Indeed, 
this term does account for the branch cut  
at $t=4 \Mp^2-\Mp^4/\Mbare^2$, which is missing in $\ghb(t)$.

\section{Infrared singularities in the self energy}\label{selfenergy}
Having established the need to sum certain graphs of the heavy baryon chiral 
perturbation series to all orders, we now formulate a general
method that leads to a representation where the relevant graphs are
automatically accounted for. The method
relies on dimensional regularization: We analyze the infrared singularities 
of the loop integrals for an arbitrary value of the dimension
$d$.

\begin{figure}[h]\centering
 \includegraphics[width=2in]{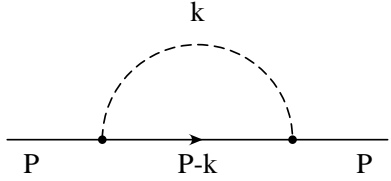}
    \caption{\label{cap:selfE} Self energy}
\end{figure}
To explain the essence of the method, we first consider the simplest example:
the self-energy graph shown in fig.~\ref{cap:selfE}.
We again focus on the
corresponding scalar loop integral
\begin{equation}\J =\frac{1}{i}\int\!\!
  \frac{d^dk}{(2\pi)^d}\,\frac{1}{(\Mp^2-k^2-i\epsilon)\,
(\Mbare^2-(P-k)^2-i\epsilon)}    \fs
\nonumber\end{equation}

\subsubsection*{\thesection.1 Singular and regular parts}\label{se1}
The integral $\J=\J(P^2,\Mp^2,\Mbare^2)$ converges for $d<4$. We 
need to
analyze it for nucleon momenta close to the mass shell: $P=\Mbare\, v + r$, 
where
$v$ is a timelike unit vector and $r$ is a small quantity of order $p$.  It is
convenient to work with the dimensionless variables
\begin{equation}\label{Omega} 
\Omega=\frac{P^2-\Mbare^2-\Mp^2}{2 \Mbare \Mp}\co\hspace{2em}
\alpha =\frac{\Mp}{\Mbare}\co\end{equation}
which represent quantities of order $\Omega=O(p^0)$ and $\alpha=O(p)$,
respectively (recall that $m$ is the larger one of the two masses,
$\alpha\simeq\frac{1}{7}$).

In the limit $\Mp\rightarrow 0$, the integral develops an infrared
singularity, generated by
small values of the variable of 
integration, $k=O(p)$. In that region, the first factor in the denominator
is of $O(p^2)$, while the second is of order $p$. Accordingly, the 
chiral expansion of $\J$ contains contributions of $O(p^{\,d-3})$. 
We may enhance
these by considering small dimensions. For $d<3$, the leading term in the 
chiral expansion of $\J$ exclusively stems from the region 
$k=O(p)$, which generates a singular contribution of order
$p^{\,d-3}$, as well as nonleading terms of order $p^{\,d-2}$, 
$p^{\,d-1}$,
$\ldots$ The remainder of the integration
region does not contain infrared singularities and thus yields a contribution 
that can be expanded in an ordinary power series. For sufficiently large,
negative values of $d$, the infrared region dominates the chiral
expansion to any desired order.

To work out the infrared singular piece, we use the standard
Schwinger-Feynman-parametrization 
\begin{equation}\label{Feynman}\frac{1}{a\, b} =\int_0^1\!\frac{dz}
{\{\,(1-z)\,a+z\, b\,\}^2}\fs\end{equation}
Performing the integration over $k$, we obtain
\begin{eqnarray}\label{J} \J\al=\al 
\kappa\!\int_0^1 \!\! dz\, C^{\,\frac{d}{2}-2}\co\hspace{2em}
\kappa=(4\pi)^{-\frac{d}{2}}\,\,\Mbare^{\,d-4}\,
\Gamma(2-\mbox{$\frac{d}{2}$})\co\\
C\al=\al z^2  -2\, \alpha\, \Omega\, z\,(1-z)
+ \alpha^2(1-z)^2-i\epsilon\fs\nonumber\end{eqnarray}
In this representation, the infrared singularity arises from small values of
$z$: There, the factor $C$ vanishes if $\alpha$ tends to zero. We may isolate
the divergent part by scaling the variable of integration, 
$z= \alpha\,u$. The upper limit
then becomes large. We extend the integration to $\infty$
and define the infrared singular part of the loop 
integral by 
\begin{eqnarray}\label{Jl} \JS \al=\al 
\kappa\!\int_0^\infty\!\! dz\, C^{\,\frac{d}{2}-2}= \kappa\,
\alpha^{d-3}\!\!\int_0^\infty\!\! du\, D^{\,\frac{d}{2}-2}\co \\
 D\al=\al 1-2\, \Omega\, u + u^2 + 2\,\alpha\,u\,(\Omega\, u-1)  + \alpha^2 
\,u^2-i\epsilon\fs\nonumber\end{eqnarray}
For short, we also refer to $I$ as the infrared part of $H$. 
The remainder is given by
\begin{equation}\label{Jh}
\JR =-\kappa\int_1^\infty\!\! dz\, C^{\,\frac{d}{2}-2}\fs\end{equation}
The decomposition\footnote{In the terminology of Ellis and Tang \cite{Tang}, 
$\JS$ represents the soft component of the amplitude, while $\JR$ is 
the hard component. We do not use these terms to avoid
confusion with the standard concepts, which concern the behaviour at
short rather than long distances.}   
\begin{equation}\label{Jlh}
\J= \JS+ \JR
\end{equation}
neatly separates the infrared singular part from the regular part: 
For noninteger 
values of the dimension, 
the chiral expansion of $\JS$ exclusively contains fractional powers of
$p$,
\bdm \JS = O(p^{\,d-3})+ O(p^{\,d-2})+O(p^{\,d-1}) + \ldots\co\edm
while the corresponding expansion of $\JR$ is an ordinary Taylor series,
\bdm \JR=O(p^{\,0}) + O(p^{\,1})+O(p^{\,2})+\ldots\edm

\subsubsection*{\thesection.2 Properties of the
  decomposition}\label{se2}

For an arbitrary value of $d$, the explicit expressions for the quantities
$\J$, $\JS$, $\JR$ involve hypergeometric functions. In four dimensions, the
corresponding integrals are elementary. We will give
the explicit representations for $d=4$ when we discuss renormalization (see
sections \ref{renormalization} and \ref{convergence}).

In terms of the dimensionless variables $\alpha$ and $\Omega$, the
chiral expansion of the infrared part takes the 
form
\bea\label{Jlexp}\JS= \Mbare^{d-4}\alpha^{d-3}\left(\rule{0em}{1em}\,
\cS^{0} +\alpha\,\cS^{1} + 
\alpha^2 \,\cS^{2} 
+\ldots\;\right)\fs\eea
The coefficients $\cS^{n}$, which only depend on $\Omega$,  
are obtained by expanding the integrand in eq.~(\ref{Jl}) in
powers of $\alpha$. The corresponding explicit expressions also
involve hypergeometric functions. 
The expansion coefficients of the
remainder, on the other hand, are simple polynomials of
$\Omega$:
\bea\label{Jhexp} \JR=\frac{m^{d-4}\,\Gamma(2-\frac{d}{2})}
{(4\pi)^{\frac{d}{2}}\,(d-3)}
\left\{1- \alpha\,\Omega+
\alpha^2\,\frac{1+(d-6)\,\Omega^2}{(d-5)}+\ldots\right\}\fs\eea
The series contains poles at $d=3,\,4,\,5\ldots\,$
We now wish to show that, for $d<3$, the chiral expansion of $\JR$ converges
throughout the low energy region.

The integrand of the representation (\ref{Jh}) is analytic in $\alpha$, except
for the cuts associated with the zeros of $C$. These are located at
\bea\alpha=\frac{z}{z-1}\,\left(-\Omega\pm\sqrt{\Omega^2-1}\right)
\fs\nonumber\eea 
The expansion of the integrand thus converges in the disk
\be\label{disk} |\alpha|< |\,\Omega\pm\sqrt{\Omega^2-1}\,|\fs\end{equation}
The condition is obeyed for all real values of  $\Omega$ in the interval
\be  
-\frac{1+\alpha^2}{2\,\alpha}<\Omega<\frac{1+\alpha^2}{2\,\alpha}
\fs \nonumber\end{equation}
This range corresponds to $0\!<P^2\!<2 \Mbare^2+2 \Mp^2$ and 
thus generously covers 
the entire low energy region. Moreover, for $d<3$, the integral converges
uniformly, so that $\JR$ is analytic in $\alpha$: The series (\ref{Jhexp}) 
converges for all values of $\Omega$ in the above interval.

In the case of $\JS$, the representation (\ref{Jl}) is relevant. 
The zeros of the term $D$ are located at
\bea \alpha=\frac{1}{u}-\Omega\pm\sqrt{\Omega^2-1}\fs\nonumber\eea
In the range $-1<\Omega<1$, the square root is imaginary, so that
the zeros occur at $|\alpha|^2=(1/u-\Omega)^2+1-\Omega^2$. This expression has
a minimum at $u=1/\Omega$. On the entire interval of integration,
the chiral expansion of the
integrand thus converges if $|\alpha|^2<1-\Omega^2$, or, equivalently
\bea\label{range2} -\sqrt{1-\alpha^2}<\Omega<\sqrt{1-\alpha^2}\fs\eea 
Again, the integral 
converges uniformly for $d<3$. Hence the 
series (\ref{Jlexp}) converges if $\Omega$ is in the range (\ref{range2}). 
This interval is considerably more narrow than the one found above, but
in the present context,
the above result suffices: It demonstrates that the two parts
occurring in the decomposition of the self energy
can unambiguously be characterized by their analytic properties at 
low energies. Since the functions $\JS$, $\JR$ are analytic
in the external momenta, their values are uniquely determined also outside
the above region. Indeed, we will show in section \ref{convergence} 
that the chiral expansion of $\JS$ 
also converges in the entire low energy region.

\subsubsection*{\thesection.3 Representation in terms of modified 
propagators}\label{se3}

It is instructive to interpret the above decomposition in terms of
the formula (\ref{Feynman}). The infrared part results if the integral
is taken from $0$ to $\infty$ rather than from $0$ to $1$,
\begin{equation}\int_0^\infty\!\frac{dz}
{\{\,(1-z)\,a+z\, b\,\}^2}=\frac{1}{a\,(b-a)}\fs\nonumber\end{equation}
This shows that the decomposition (\ref{Jlh}) corresponds to the two terms in 
the algebraic identity
\begin{equation}\label{ab}\frac{1}{a\,b}=\frac{1}{a\,(b-a)}-
\frac{1}{b\,(b-a)}\fs
\end{equation} 
Note, however, that the $i\epsilon$-prescription drops out in the difference 
$b-a$, so that the integral over $k$ of the individual terms on the right 
is ambiguous. To avoid the ambiguity, we may for instance equip the two
masses with different imaginary parts. The proper
expression for the infrared part reads
\begin{equation}\label{Jlint}\JS =\frac{1}{i}\int\!\!
  \frac{d^dk}{(2\pi)^d}\,\frac{1}{(\Mp^2-k^2-i\epsilon)\,
(\Mbare^2-P^2 + 2 Pk -\Mp^2-i\epsilon)} 
   \fs \end{equation}
It differs from $\J$ in that the term $k^2$ occurring in the nucleon
propagator is replaced by $\Mp^2$. The regular
part, on the other hand, is given by
\begin{equation}\label{Jhint}\JR =-\frac{1}{i}\int\!\!
\frac{d^dk}{(2\pi)^d}\,\frac{1}{(\Mbare^2-(P-k)^2-i\epsilon)\,
(\Mbare^2-P^2 + 2 Pk -\Mp^2-i\epsilon)} 
   \fs
\end{equation}
The expression involves two heavy particle propagators, one where the term
with $k^2$ is retained, one where this term is replaced by $\Mp^2$. 

The function $\J$ is symmetric with respect to an interchange of $\Mp$ and
$\Mbare$. The term $\JS$ collects those contributions of the expansion in
powers 
of $\alpha=\Mp/\Mbare$ that involve fractional powers of $\Mp$, while $\JR$
collects the fractional powers of $\Mbare$.  The operation $\Mp\leftrightarrow
\Mbare$, $k\leftrightarrow P-k$ interchanges $a$ and $b$, so that the two terms
on the r.h.s.~of eq.~(\ref{ab}) are mapped into one another.  
This suggests that
an interchange of the two masses takes $\JS$ into $\JR$ and vice versa. The
argument does not go through, however: The difference in the
$i\epsilon$-prescriptions shows that the operation does not take the integral
in eq.~(\ref{Jlint}) into the one in eq.~(\ref{Jhint}). The sum $\JS+\JR$, of
course, is symmetric under an interchange of the two masses.

In the heavy baryon chiral perturbation series, the scalar self energy graph 
of fig.~\ref{cap:selfE} is replaced by an infinite string of one loop graphs,
\begin{figure}[h]\centering
\includegraphics[width=5.5in]{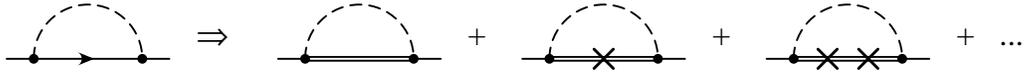}
    \caption{\label{cap:insertions} Internal line insertions. The double line
      denotes the heavy baryon propagator 
$[2\Mbare\, v\cdot( k-r)-i\epsilon]^{-1}$, the
      cross an insertion of $(k-r)^2$.}
\end{figure}
involving an arbitrary number of internal line insertions
(fig.~\ref{cap:insertions}).
The corresponding scalar loop integrals are
obtained from the relativistic version with $P=\Mbare\,v + r$, treating
both $r$ and $k$ as small quantities of order $p$ and expanding the nucleon
propagator in powers of $p$:
\bea\frac{1}{2 \Mbare\, v\cdot(
  k-r) -(k-r)^2-i\epsilon} \al=\al \nonumber\\\al\al\hspace{-6em} \frac{1}{2
  \Mbare\, v\cdot( k-r)-i\epsilon} +\frac{(k-r)^2}{\{2 \Mbare\, v\cdot(
  k-r)-i\epsilon\}^2}+ \ldots\nonumber 
\eea 
The integral over the leading term
converges for $d<3$, yielding a contribution of order $p^{\,d-3}$ that depends
on the vector $r$ only through the projection $v\cdot r$. The integral over
the second term of the series converges for $d<2$ and yields a term of order
$p^{\,d-2}$, etc. The individual contributions are not Lorentz invariant, but
the series may be reordered in such a manner that only the Lorentz invariant
combination $ 2\Mbare\,v\cdot r +r^2=P^2-\Mbare^2 $ enters: The heavy baryon 
series
then reproduces the expansion of $\JS$ in eq.~(\ref{Jlexp}), term by term.

To demonstrate that this is so, we recall that the infrared part
dominates the chiral expansion to any desired order if $d$ is taken
sufficiently negative: For $d< 3-n$, the region $k=O(p)$ yields all of the
terms in the chiral expansion of the integral $\J$, up to and including
$p^{\,d-3+n}$. In that region of integration, however, it is legitimate to
interchange the integration with the expansion. This is precisely what is done
in the heavy baryon approach. Hence that approach does yield the expansion of
$\JS$, to any finite order: The infrared part of the relativistic
loop integral represents the sum of the corresponding integrals occurring in
the heavy baryon series.  The difference between the two formulations of the
effective theory resides in the regular part: In the heavy baryon approach,
this part is absent. The representation (\ref{Jhint}) shows that $\JR$
corresponds to a loop formed with two nucleon lines -- evaluating integrals of
this type in the manner described above, the rules of HB\ch PT \cite{rules}
indeed yield $\JR=0$, order by order.

\subsubsection*{\thesection.4 Dispersive representation}\label{se4}
The function $\J(P^2,\Mp^2,\Mbare^2)$ obeys a dispersion
relation, which in four dimensions requires one subtraction. Setting $s=P^2$
and suppressing the other variables, the relation takes the form
\bea\label{dispJ} \J(s)\al=\al\J(s_0)+\frac{s-s_0}{\pi}\!  \int_{\sth}^\infty
\frac{ds'}{(s'-s_0)\,(s'-s-i\epsilon)}\;\mbox{Im}\,\J(s')\co\\
\mbox{Im}\,\J(s)\al=\al\frac{\;\rho(s)}{16\,\pi\,s}\;
\theta(s-s_+)\co\hspace{2em}s_{\pm}=(\Mbare\pm \Mp)^2\fs\nonumber\eea The
function $\rho(s)$ stands for the familiar two-body phase space factor \bdm
\rho(s)=\sqrt{(s-s_+)\,(s-s_-)}= 2\,\Mp \Mbare\sqrt{\Omega^2-1}\fs\edm The
logarithmic divergence of the loop integral manifests itself in the
subtraction constant $\J(s_0)$, which contains a pole at $d=4$. If we subtract
at threshold, $s_0=s_+$, the subtraction constant is given by\footnote{While
  the representation for the subtraction term holds in any dimension, the one
  for the discontinuity is valid only in the limit $d\rightarrow 4$.}
\bdm\J(s_+)=\frac{\Gamma(2-\frac{d}{2})} {(4\pi)^\frac{d}{2}(d-3)}\;
\frac{\Mbare^{d-3}+ \Mp^{d-3}}{(\Mbare+\Mp)}\fs\edm The structure of this
expression is typical: It contains a term with a fractional power of $\Mp$ and
one with a fractional power of $\Mbare$, representing the infrared singular
and regular parts, respectively,
\bea\JS(s_+)\al=\al\frac{\Gamma(2-\frac{d}{2})} {(4\pi)^\frac{d}{2}(d-3)}\;
\frac{ \Mp^{d-3}}{(\Mbare+\Mp)}\co\nonumber\\
\JR(s_+)\al=\al\frac{\Gamma(2-\frac{d}{2})} {(4\pi)^\frac{d}{2}(d-3)}\;
\frac{\Mbare^{d-3}}{(\Mbare+\Mp)}\fs\nonumber\eea
 
The dispersion relation (\ref{dispJ}) is a variant of the formula (\ref{J}):
In the limit $d\rightarrow 4$, the corresponding representation for 
$\J(s)-\J(s_+)$ is proportional to $\int_0^1 dz\,\mbox{ln}(C/C_+)$, 
where $C_+$ is the value of $C$ at $s=s_+$. Since $C$ is linear in $s$, 
\bea C\al=\al \frac{z(1-z)}{\Mbare^2}\;(s'-s-i\epsilon)\co\nonumber\\
s'\al=\al\frac{\Mbare^2}{1-z}+\frac{\Mp^2}{z}\co\hspace{2em} 
\frac{ds'}{\rho(s')}=\frac{|dz|}{|z(1-z)|}\co\nonumber\eea
the argument of the logarithm reduces to $(s'-s-i\epsilon)/(s'-s_+)$. With an 
integration by parts this indeed leads to eq.~(\ref{dispJ}). 
Note that, on the interval $0<z<1$, 
the function $s'=s'(z)$ has a minimum at
\bdm z_{min}=\frac{\Mp}{\Mbare+\Mp}\co\hspace{2em}
s'_{min}=s_+\fs\edm 
The map $z\rightarrow s'$ thus covers the interval
$(\Mbare+\Mp)^2<s'<\infty$ twice. 

In the Feynman parameter representation, the only difference between
$\J$,  $\JS$ and $\JR$ is that the integrations extend over 
different intervals.
The one relevant for $\JR$ is $1<z<\infty$, which is mapped onto 
$-\infty<s'<0$. Accordingly, $\JR$ has a
cut along the negative real axis, but is analytic in the right half plane.
The discontinuity across the cut is only half as big as in the case of
$\J$, because the interval is now covered only once:
\be \mbox{Im}\,\JR(s)=\frac{\;\rho(s)}
{32\,\pi\,s}\;\theta (-s)\fs\nonumber\end{equation}
The expression only holds for $s\neq 0$: In addition to the cut, the 
function $\JR(s)$ contains a pole at $s=0$. 
In view of the singular behaviour of
the discontinuity, the dispersion relation 
cannot be written in the form (\ref{dispJ}). Instead, we may establish a twice 
subtracted dispersion relation for $s\,\JR(s)$ (compare eq.~(\ref{Jpole}) in
section \ref{renormalization}):
\bea \JR(s)\al=\al \frac{s+\Mbare^2-\Mp^2}{s}\;R_0
+\frac{(s-s_+)\,(s-s_-)}{32\,\pi^2\,s}\!
\int_{-\infty}^{\,0}\frac{ds'}{\rho(s')\,(s'-s-i\epsilon)}\co\nonumber\\
R_0\al=\al \frac{\Mbare+\Mp}{2\Mbare}\;\JR(s_+)
\fs\nonumber\eea
In contrast to the subtraction term in the dispersion relation for the full
integral $\J(s)$, the one occurring here does not represent a constant,
but contains a pole at $s=0$. That point, however, is far outside the 
region covered by chiral perturbation theory. Neither the subtraction term
nor the dispersion integral contain singularities in the low energy region.
The dispersive representation neatly demonstrates that the chiral 
expansion of $\JR$ is an ordinary Taylor series.

The function $\JS$ is the difference between $\J$ and $\JR$ and 
hence has a cut on the left as well as one on the right:
\be\mbox{Im}\,\JS(s)=\frac{\;\rho(s)}{16\,\pi\,s}\;
\left\{\theta(s-s_+)-\frac{1}{2}\,\theta (-s)\right\}\fs\nonumber\end{equation}
Note that
the infrared part possesses the same discontinuity across the
right hand cut as the full integral, even far away from threshold.

\section{Generalization}
\label{generalization}

We now generalize the above analysis to arbitrary one loop graphs.
All of these can be reduced to integrals of the form 
\begin{equation}
\J_{mn}^{\mu_1\ldots\mu_r}=\frac{1}{i}\int\!\!
  \frac{d^dk}{(2\pi)^d}\,\frac{k^{\mu_1}\ldots k^{\mu_r}}
{a_1\ldots a_m\,b_1\ldots b_n}    \fs
\nonumber\end{equation}
The denominator stems from the meson and 
nucleon propagators: 
\begin{eqnarray}a_i\al=\al \Mp^2-(k-q_i)^2-i\epsilon\co\nonumber\\
b_j\al=\al \Mbare^2-(P_j-k)^2-i\epsilon\fs\nonumber\end{eqnarray}
Part of the numerator is generated by the derivative couplings 
characteristic of chiral perturbation theory. The remainder arises 
from the term $k\hspace{-0.5em}\slash$ that occurs in the numerator of the
nucleon propagator, $P_j\hspace{-0.9em}\slash\;\,-k\hspace{-0.5em}\slash\,+m$. 
The external meson momenta $q_i^\mu$ represent quantities of order $p$.
The nucleon momenta $P_j^\mu$ are close to the mass shell, 
$P_j^2=\Mbare^2+O(p)$.

On account of Lorentz invariance, the above integral 
may be decomposed in a basis formed with tensor polynomials of the external
momenta $P_i^\mu$, $q_k^\mu$. Inverting this representation, the coefficients 
of the decomposition may be expressed in terms of scalar integrals, where
the momentum factors are replaced by their projections onto the 
external momenta or by factors of $k^2$. These, however, represent 
linear combinations of the terms $a_i$, $b_k$ occurring in
the denominator:\footnote{Without loss of generality, we may put one of the
external momenta to zero, say $q_1$.} \bdm
k^2=\Mp^2-a_1\co\hspace{1em} P_1\cdot 
k=\mbox{$\frac{1}{2}$}\,(b_1-a_1+P_1^2-\Mbare^2+\Mp^2)
\co\hspace{1em}\ldots\edm 
If the graph in question involves several different external momenta, 
the above procedure may be rather clumsy, but it shows that
all one loop integrals arising in the 
$\pi N$-system may be expressed in 
terms of the scalar functions 
\begin{equation}\label{Jmn}
\J_{mn}=\frac{1}{i}\int\!\!
  \frac{d^dk}{(2\pi)^d}\,\frac{1}
{a_1\ldots a_m\,b_1\ldots b_n}    \co
\end{equation}
so that it suffices to study the
properties of these.\footnote{A more efficient method is described in
ref.~\cite{Tarasov}. As shown there,
the tensor integrals may be generated by applying suitable differential 
operators to the scalar ones.}

\subsubsection*{\thesection.1 Singular and regular parts}\label{gen1}

The self energy corresponds to $\J=\J_{11}$, the triangle graph to 
$\gamma=\J_{21}$. Concerning the counting of powers of momentum, loops 
that do not involve the propagation of a heavy particle are trivial: The 
integral then represents a homogeneous function of order $p^{\,d-2m}$, so that
a regular part does not occur,
\bea \JS_{m 0}=\J_{m 0}\co\hspace{2em}\JR_{m 0}=0\fs\nonumber\eea

In the opposite extreme, where the graph exclusively involves nucleon
propagators, we may shift the variable of integration with $k'=k-P_1$.  The
denominator then depends on the external momenta only through the differences
$P_j-P_1$, which represent small terms of $O(p)$.  In Euclidean space, the
integrand is thus approximately given by $(m^2+k^{\prime\,2})^{-n}$, so that
the region $k'=O(m)$ dominates.  For $d<2n-r$, the first $r$ terms of the
chiral expansion may be worked out by performing the expansion under the
integral -- the coefficients are polynomials in the external momenta. Hence
the integral does not contain any infrared singularities, \bea
\JS_{0 n}=0\co\hspace{2em}\JR_{0 n}=\J_{0 n}\fs\nonumber\eea

If the loop contains meson as well as nucleon propagators, the integral 
involves both an infrared singular and a regular piece. 
The infrared singularities arise from the region $k=O(p)$.
There, each of the
pion propagators yields a factor of $p^{-2} $, while each of the nucleon
propagators yields a factor of $p^{-1}$, so that the infrared region
generates a contribution of order $p^{\,d-2m-n}$. 

We may combine all of the mesonic propagators by means of the formula
\begin{equation}\frac{1}{a_1\ldots a_m}=
\left(-\frac{\partial}{\partial \Mp^2}\right)^{\!\!(m-1)}\!\!
\int_0^1\!\! dx_1\ldots\int_0^1\!\! dx_{m-1}\;\frac{\,X}{A}\fs
\nonumber\end{equation}
The numerator is given by ($m\geq 2$ and $X=1$ if $m=2$)
\be
X= x_2\cdot (x_3)^{\,2} \cdots  (x_{m-1})^{m-2}\fs
\nonumber\end{equation}
The denominator is obtained recursively, with
\bea A_{p+1}\al =\al x_p\, A_{p} +(1-x_p)\, a_{p+1}\hspace{3em} 
(p=1,\ldots\,,m-1)\nonumber\\
A_1 \al =\al a_1 \co \hspace{2em}A=A_{m}\fs\nonumber\eea
The result for $A$ is quadratic in $k$,
\begin{equation}A=\bar{A}  - (k-\bar{q})^2 -i\epsilon\fs 
\nonumber\end{equation}
The constant term $\bar{A}$ is of order $p^2$, while $\bar{q}$ represents a 
linear combination of external momenta and is of order $p$.
Likewise, if there are several nucleon propagators, we may combine these with
\begin{equation}\frac{1}{b_1\ldots b_n}=
\left(-\frac{\partial}{\partial \Mbare^2}\right)^{\!\!(n-1)}\!\!\!
\int_0^1\!\! dy_1\ldots\int_0^1\!\! dy_{n-1}\;\frac{\,Y}{B}\co\hspace{0.7em}
Y= y_2\cdot (y_3)^{\,2} \cdots  (y_{n-1})^{n-2}\,.
\nonumber\end{equation}
In this case, the denominator is of the form
\begin{equation}B=\bar{B} -(\bar{P}-k)^2-i\epsilon \co\nonumber \end{equation}
with $\bar{P}^2=\Mbare^2 + O(p)$, $\bar{B}=\Mbare^2+O(p)$.
The loop integral then becomes
\begin{equation}\J_{mn}=
\left(-\frac{\partial}
{\partial \Mp^2}\right)^{\!\!(m-1)}\!\!
\left(-\frac{\partial}{\partial \Mbare^2}\right)^{\!\!(n-1)}\!\!
\int_0^1\!\! dxdy\,X Y\;\frac{1}{i}\int\!\!
  \frac{d^dk}{(2\pi)^d}\,
\frac{1}{A\, B}    \co
\nonumber\end{equation}
where we have abbreviated the integral over the $m+n-2$ Feynman parameters
by $\int_0^1 dxdy$. 
The integral over $k$ is the one studied in section \ref{selfenergy},
\begin{equation}\frac{1}{i}\int\!\!
  \frac{d^dk}{(2\pi)^d}\,\frac{1}{A\, B}=
\J(P^2,\Mp^2,\Mbare^2)\;\rule[-0.5em]{0.03em}{1.3em}_{\;P\rightarrow 
\bar{P}-\bar{q},\;\Mp^2\rightarrow \bar{A} ,\; \Mbare^2\rightarrow
\bar{B}}\hspace{1em}\co\nonumber   
\end{equation}
so that the analysis given there may be taken over. The 
decomposition $\J=\JS+\JR$ leads to an analogous
splitting for the general scalar one loop integral:
\be \J_{m n}=\JS_{mn}+\JR_{mn}\fs\nonumber\end{equation}
In particular, the representation for $\JS_{mn}$ reads  
\begin{equation}\label{phiJS}\JS_{mn}=
\int_0^1\!\! dxdy\,X Y
\left(-\frac{\partial}{\partial \bar{A}}\right)^{\!\!(m-1)}\!\!
\left(-\frac{\partial}{\partial \bar{B}}\right)^{\!\!(n-1)}\!\!
  \JS\left(\rule{0em}{0.7em}
(\bar{P}-\bar{q})^2,\bar{A},\bar{B}\right)    \fs
\end{equation}
In view of the formula (\ref{Jl}), the representation of the 
infrared part in terms of Feynman parameters coincides with the one 
obtained for the full integral $\J_{mn}$, except that
the integration over one of these parameters -- 
the one needed to combine the meson propagators with the nucleon propagators
-- runs from $0$ to $\infty$ rather than from $0$ to $1$. Performing the
derivatives with respect to the masses and integrating over $k$, 
we obtain
\bea\label{phimnS} \JS_{mn}\al =\al \frac{\Gamma(m + n-
\mbox{$\frac{d}{2}$})}{ (4\pi)^{\frac{d}{2}}}
\int_0^1\!\! dxdy\,X Y \!\int_0^\infty\!\!\!dz\, (1-z)^{m-1}\,z^{n-1} \;
C^{\,\frac{d}{2}-m-n}
\co\nonumber\\
C\al =\al (1-z)\,\bar{A}+z\,\bar{B} - z\,(1-z)\,(\bar{P}-\bar{q})^2
-i\epsilon \fs 
\eea
The analogous representation for $\JR_{mn}$ is obtained by replacing the
interval of integration for $z$ by $1<z<\infty$ and changing the overall sign. 
\subsubsection*{\thesection.2 Uniqueness of the decomposition}\label{gen2}

Formally, we may expand this representation in the same manner as
the self energy: Set $P_j= \Mbare\, v + r_j$, rescale the variable of 
integration with $z =\alpha\,u$ 
and expand the integrand in powers of $q_i,\,r_j$ and $\Mp$. The same result
is obtained, term by term, if we treat the loop momentum $k$ in the
integral (\ref{Jmn}) as a quantity
of $O(p)$ and perform the chiral expansion under the integral. Applying 
the procedure also to $\JR_{mn}$, we obtain two 
series of the form
\bea\label{chexp} \JS_{mn}\al=\al
\Mbare^{d-2m-2n}\alpha^{d-2m-n}\left(\rule{0em}{1em}\,\cS_{mn}^{0} +
\alpha\,\cS_{mn}^{1} + 
\alpha^2\, \cS_{mn}^{2} +\ldots\;\right)\co\\
\JR_{mn}\al=\al \Mbare^{d-2m-2n}\left(\rule{0em}{1em}\,
r_{mn}^0 + \alpha\,r_{mn}^1+
\alpha^2\,r^2_{mn}+\ldots\;\right)\fs\nonumber\eea
While the dimensionless coefficients $\cS^{p}_{mn}$ are nontrivial 
functions of the variables $q_1/\Mp,\,\ldots\,,q_m/\Mp $ and 
$r_1/\Mp,\,\ldots\,, r_n/\Mp$, those occurring in the expansion of $\JR_{mn}$
are polynomials. 

As discussed in detail in section \ref{scalar}, the chiral expansion in general
makes sense only in part of the low energy region. In the representation
(\ref{phimnS}), the problem arises from the fact that it is not always
legitimate to interchange the integration over the Feynman parameters with
this expansion -- the integral $\JS_{mn}$ as such describes the low energy 
behaviour perfectly well. 

We again emphasize that the decomposition of the integral into 
an infrared part and a remainder is uniquely characterized by the analytic
properties of the two pieces, also if
the expansion (\ref{chexp}) only holds in part of the low energy region. 
The proof closely follows the one
given in section 5.2 and we only indicate the modifications needed to
adapt it to the present more general situation. 
As shown there, the chiral expansion of $\JS$ and $\JR$
does converge in part of the low energy region. In the present context,
the range (\ref{range2}) corresponds to
\bea\label{rangePq} -2\left\{\rule{0em}{0.9em}\bar{A}\,
(\bar{B}-\bar{A})\right\}^\frac{1}{2}<
(\bar{P}-\bar{q})^2-\bar{A}-\bar{B}< 
2\left\{\rule{0em}{0.9em}\bar{A}\,(\bar{B}-\bar{A})\right\}^\frac{1}{2}
\fs\eea
The quantities $\bar{A}$, $\bar{B}$, $\bar{q}$ and $\bar{P}$ 
depend on the parameters $x_1,\,\ldots\,, x_{m-1}$,
$y_1,\,\ldots\,,y_{n-1}$ used to combine the meson and nucleon propagators,
respectively -- the condition (\ref{rangePq}) must be met for all values of 
these parameters
in the range $0\leq x_i, y_j\leq 1 $. If this is the case, the
chiral expansion of the integrand on the r.h.s.~of eq.~(\ref{phiJS}) converges.

It is not difficult to see that the low energy region does contain a domain
where the above condition is met. The vector $\bar{q}$ is
contained in the polyhedron spanned by the corners $q_i$ and
analogously for $\bar{P}$. We may, for instance, take all of the $q_i$ to
be much smaller than $M$ and put the $P_i$ in the immediate vicinity of
a vector $P$ that sits on the nucleon mass shell, such that $\bar{A}\simeq
M^2$, $\bar{B}\simeq m^2$, $(\bar{P}-\bar{q})^2\simeq m^2$. 
The condition is then obviously met in the entire region of integration. 
Moreover, the integral over the Feynman parameters converges uniformly there. 
This completes the proof.

\section{Comparison with \bHBCHPT }
\label{comparison}

In the framework of the relativistic effective theory, the evaluation of an
amplitude to one loop thus yields three categories of contributions, arising
from (a) tree graphs, (b) infrared singular part and (c) regular part of the
one loop integrals. At a given order of the chiral expansion, the one particle
irreducible components of the regular parts are polynomials in the external
momenta. In coordinate space, these contributions thus represent local terms:
They are equivalent to the tree graph contributions generated by a suitable
Lagrangian, $\Delta {\cal L}$. So, if we replace the effective Lagrangian by
\be {\cal L}^{\prime}_\eff={\cal L}_\eff +\Delta {\cal L}\co
\nonumber\end{equation} 
we may drop the regular parts of the loop integrals.
The resulting representation is identical to the one obtained in the heavy
baryon approach, except that the infrared parts of the one loop graphs are
included to all orders -- the problems afflicting the chiral expansion of the
infrared singularities are avoided.  We add a few remarks concerning the above
relation between the term ${\cal L}_\eff$ relevant for the original form of
the effective theory and the effective Lagrangian occurring in our framework,
${\cal L}^{\prime}_\eff$.

\subsubsection*{\thesection.1 Lorentz invariance}\label{hb1}
First, we note that both of these schemes are characterized by a Lorentz
invariant effective
Lagrangian: By construction, the term $\Delta{\cal L}$ is Lorentz invariant.
In explicit formulations of HB\ch PT, the invariance of
the effective Lagrangian is by no means manifest.
One of the reasons is that
the equations of motion are used to eliminate two of the four components of the
Dirac spinor that describes the nucleon in the relativistic formulation of 
the theory.
In fact, it is perfectly legitimate to use the equations of motion: The
resulting modification of the effective Lagrangian is equivalent to a 
change of variables. The operation, however, destroys manifest
Lorentz invariance -- in terms of the new variables, the transformation
law of the field takes a rather complicated form. 

The point here is
that all of that can be avoided. Instead of explicitly performing the
chiral expansion of the Lagrangian and evaluating the perturbation series
with the corresponding nonrelativistic propagators, we may simply replace the
integrands of the various loop integrals by their chiral series,
i.e.~perform the nonrelativistic expansion before doing the integral
\cite{Tang}. As discussed above, this
procedure amounts to replacing the relativistic loop integrals by their 
infrared parts. The 
result for the various amplitudes of interest is the same as the one
obtained within the standard approach, except that our method accounts for
the mass insertions to all orders. 

\subsubsection*{\thesection.2 Chiral symmetry}\label{hb2}
Both the relativistic and the heavy baryon formulations of the effective
theory are based on an effective Lagrangian that is manifestly invariant
under chiral transformations. In the above construction, this property of
the term ${\cal L}^{\prime}_\eff$ is not evident. Although the equivalence
with the standard heavy baryon approach ensures chiral symmetry, it
is instructive to see how this property arises within the present framework.

For this purpose, we first 
formulate chiral symmetry in terms of
objects that are amenable to an evaluation in the framework of the effective
theory: Consider all Green functions of the 
type $\langle N(P',s')|T\,O_1\ldots O_r|N(P,s)\rangle$, 
where the operators $O_i$ represent vector, axial,
scalar or pseudoscalar quark currents. Chiral symmetry implies that,
in the limit where the quark masses are put equal to zero, these matrix 
elements are interrelated through a set of Ward identities. We now analyze the
implications of these identities for the regular parts of the one loop graphs
responsible for the term $\Delta{\cal L}$.

At one loop, the Green functions
are represented by a sum of three contributions belonging to the three 
categories a, b, c specified above.
Since ${\cal L}_\eff$ is invariant under chiral symmetry, the tree
graph contributions (a) obey the Ward identities. 
Furthermore,
dimensional regularization preserves the symmetries of the Lagrangian.
Hence the contributions from the one loop graphs (b + c) also obey these
identities, for any value of the regularization parameter $d$. Now, the chiral
expansion of the infrared singular (regular) part only contains fractional 
(integer) powers of the chiral expansion parameter $p$. 
Hence the Ward identities can be satisfied by the sum
of the two pieces only if they are obeyed separately by the two
parts: The vertices collected in $\Delta {\cal L}$
obey the same set of linear constraints as the vertices contained in
${\cal L}_\eff$. This explains why the term $\Delta {\cal L}$ is invariant
under chiral transformations\footnote{For the case of the purely mesonic 
vertices, it is explicitly demonstrated 
in ref.~\cite{foundations} that the Ward identities indeed imply a symmetric
effective Lagrangian, but we did not perform the corresponding analysis for 
the present, more general case.}.

\subsubsection*{\thesection.3 Infrared part as an alternative 
regularization}\label{hb3}
Since ${\cal L}_\eff$ contains all terms permitted by Lorentz
invariance and chiral symmetry, the modification 
${\cal L}_\eff\rightarrow{\cal L}_\eff+\Delta {\cal L}$ is equivalent
to a change of the effective coupling constants: 
The bare coupling constants to be used
in the original form of the relativistic effective theory differ from those
occurring in our scheme.
In this respect, the two methods of calculation
appear like two different regularizations of the theory -- for the physical
amplitudes to be independent thereof, the values of the 
bare couplings must be tuned to the regularization used. 

Note, however,
that the two prescriptions for the evaluation of the loop integrals
in general lead to different results even if these are convergent.
Viewing $\J_{mn}^{\mu_1\ldots\mu_r}$ and $\JS_{mn}^{\mu_1\ldots\mu_r}$ 
as two different 
regularizations of the same integral, we are leaving the standard class 
of admissible regularizations (dimensional, 
Pauli-Villars, momentum space cutoff, $\ldots\;$). 

For a Lagrangian that contains {\it all} of the Lorentz invariant 
vertices that can be formed with the field and its derivatives, 
it would be natural to take the loop integrals as being defined only 
modulo an arbitrary Lorentz invariant polynomial.
In our context, that class is too large. Since the effective
Lagrangian is chirally invariant, the same coupling constant
determines the strength of an entire string of vertices. A change in 
one of the couplings generates a specific polynomial contribution in 
several different amplitudes. Conversely, if we are removing a polynomial
from one of the loop integrals, we need to subtract a corresponding term
in some of the other loop integrals, too -- otherwise, the procedure would 
yield amplitudes that do not obey all of the Ward
identities. It is essential here that dimensional
regularization preserves the symmetry and that this procedure
allows us to unambiguously identify the infrared part for all of the 
integrals. At any finite order of the chiral expansion, the difference
between the dimensional regularization $H_{mn}^{\mu_1\ldots\mu_r}$ 
and the infrared
part $\JS_{mn}^{\mu_1\ldots\mu_r}$ is a polynomial and the polynomials 
occurring in
different loop integrals are correlated in such a manner that the Ward
identities are obeyed:
If we consistently replace all of the integrals by 
their infrared parts, the content of the theory
remains the same. Hence, it is legitimate to think of the infrared part
as an alternative regularization of the loop integrals and to indicate this
with the symbol $\int_Id^dk$ :
\bea 
\JS_{mn}^{\mu_1\ldots\mu_r}=\frac{1}{i}\int_I
  \frac{d^dk}{(2\pi)^d}\,\frac{k^{\mu_1}\ldots k^{\mu_r}}
{a_1\ldots a_m\,b_1\ldots b_n}    \fs\nonumber\eea

The standard regularizations yield the smoothest possible high energy
behaviour: The expansion in inverse powers of $p$ starts with
$p^{\,d+r-2m-2n}$. The high
energy behaviour is crucial for renormalizability but in the context of
effective low energy theories, it is irrelevant, because the high energy
domain is anyway outside the reach of the framework.  The infrared
regularization $\JS_{mn}^{\mu_1\ldots\mu_r}$ instead yields maximally smooth
behaviour at low energies: All of the regular contributions of order
$p^0,\,p^1,\,\ldots$ are absorbed in the effective coupling constants, so that
the low energy expansion starts with an infrared singular piece of order
$p^{\,d+r-2m-n}$. 

In principle, the analysis given in the preceding section may be extended to
arbitrary graphs. The leading infrared singularity originates in the region
where all of the loop momenta are small. Disregarding momentum factors that
may arise from the vertices or from the propagators, the leading infrared
singularity occurring in the low energy expansion of a graph with $\ell$
loops, $m$ mesonic and $n$ baryonic propagators is of order $p^{\,\ell\,
  d-2m-n}$ -- the counting of powers is the same as in heavy baryon chiral
perturbation theory.  At the present stage of our understanding, the extension
beyond the one loop approximation is a rather academic issue, however.  In the
case of $\pi$N scattering for instance, significant progress could be achieved
by extending the known results to order $p^4$. With the method outlined above,
this should require rather little effort: It suffices to (i) replace the
dimensional regularization used in ref.~\cite{gss} by the infrared
regularization and (ii) add the contributions from the one loop graphs
generated by ${\cal L}_2$.

\section{Renormalization}
\label{renormalization}
We now consider the behaviour of the loop integrals in the
limit $d\rightarrow 4$ and again start with the self
energy. 

\subsubsection*{\thesection.1 Self energy}\label{ren1}
The integral 
over the Feynman parameter $z$ in eq.~(\ref{Jl}) only converges for
$d<3$. The continuation to $d\rightarrow 4$ may be performed as follows.
The factor $C$ is of the form
\bea C\al =\al C_0+ C_1(z-z_0)^2\co\hspace{2em}
z_0=\frac{\alpha\,(\Omega + \alpha)}{1+2\, \alpha\, \Omega + \alpha^2}\co
\nonumber\\
C_0\al=\al\frac{\alpha^2 (1-\Omega^2)}{1+2\, \alpha\, \Omega + \alpha^2}
\co\hspace{3.1em}
C_1=(1+2\, \alpha\, \Omega + \alpha^2)\fs\nonumber\eea 
Replace $C^{\frac{d}{2}-2}$ by $ C_0\,C^{\frac{d}{2}-3}$+ $C_1(z-z_0)^2 
\,C^{\frac{d}{2}-3}$. The second term is proportional to the derivative
of $C^{\frac{d}{2}-2}$. An integration by parts leads to
\be\label{Jpole} \JS=\kappa \int_0^\infty 
\!\!dz\, C^{\frac{d}{2}-2}=
\frac{\kappa }{d-3}\left\{ \alpha^{d-4}\,z_0  +
(d-4)\,C_0\int_0^\infty
\!\!dz\, C^{\frac{d}{2}-3}\right\}\fs\end{equation}
Since the remaining integral converges for $d<5$, the right hand side 
can now be 
continued analytically to $d=4$. The factor $\kappa$ contains a pole there,
\bea \kappa\al=\al\frac{\Gamma(2-\mbox{$\frac{d}{2}$})}
{(4\pi)^{\frac{d}{2}}}\,\Mbare^{d-4}
=- 2\,\lambdabar 
-\frac{1}{16\pi^2}+O(d-4)\co\nonumber\\
\lambdabar\al=\al\frac{m^{d-4}}{(4\pi)^2}\left\{\frac{1}{d-4}-\frac{1}{2}
\left(\rule{0em}{1em}\,\mbox{ln}\, 4\pi +\Gamma'(1)+1\right)\right\}
\fs\nonumber\eea
We have expressed the singularity in terms of the standard pole term 
$\lambda$, which contains a running scale $\mu$. The scale relevant here
is the mass of the nucleon: $\lambdabar$ represents the value of $\lambda$ at
the scale $\mu=m$.
The renormalized amplitude, which we denote by $\JSbar$, is obtained by 
removing the pole   
\bea \JSbar \al=\al \JS-\lambdabar\,\Jdiv\co \nonumber\\
\Jdiv\al=\al-\frac{P^2-\Mbare^2+\Mp^2}{P^2}\fs\nonumber\eea
For the regular part and for the full integral, the renormalizations 
read
\be \JRbar= \JR +(2+\Jdiv)\,\lambdabar\co\hspace{2em}
\Jbar=\J + 2\, \lambdabar\fs\nonumber\end{equation}
By construction, we have
\bdm \Jbar=\JSbar+\JRbar\fs\edm
The counter term for $\J$ is momentum independent, but the
quantity $\Jdiv$ is not. As $\nu$ does not contain infrared
singularities, its expansion in powers of $M^2$ and $P^2-m^2$ 
is an ordinary Taylor series.

\subsubsection*{\thesection.2 Renormalization scale}\label{ren2}
It is important here that, in the relativistic formulation of the effective 
theory, loops involving
nucleon propagators contain an intrinsic scale, even in the chiral limit:  
the nucleon mass. This is in marked contrast to the mesonic sector 
and to the standard heavy baryon
approach, where dimensionally regularized loop integrals are scale 
invariant in the chiral limit,
so that the removal of the divergences necessarily 
involves a free parameter, the
running scale. As is well known, this does not give rise to ambiguities,
because 
the renormalization of the loop integrals only requires polynomial counter
terms: It suffices to tune the 
coupling constants of those
terms in the effective Lagrangian that enter at the order of the chiral
expansion considered -- the result for
quantities of physical interest then becomes scale independent. 

In the present context, the situation is different. The chiral expansion of
the infrared part $\JS$ contains arbitrarily high orders -- in the language of
standard HB\ch PT, we are summing up an infinite number of graphs. Their
renormalization requires counter term polynomials of arbitrarily high order:
Heavy baryon graphs of $O(p^n)$ call for counter terms of $O(p^n)$. Indeed,
the chiral expansion of the counter term $\bar{\lambda} \,\nu$ contains 
polynomials
of arbitrarily high order.  If we were to work with an arbitrary running
scale, we would need to include infinitely many terms in the effective
Lagrangian and tune their scale dependence properly -- only then the
amplitudes would become scale independent.

Neither can this be done in practice, nor is it necessary.
All of the loop integrals that require counter terms with a nonpolynomial
momentum dependence contain an intrinsic scale and
we may identify the renormalization scale with this one, i.e.~set $\mu=m$. 
A running scale is needed only 
for loops formed exclusively with mesonic propagators -- these do not contain
an intrinsic one. 

\subsubsection*{\thesection.3 Other one loop graphs}\label{ren3}

The generalization to other one loop graphs is straightforward.
Concerning the full
scalar integrals, only the self energy $\J$ requires renormalization: 
For $m+n>2$ the functions $\J_{mn}$ represent convergent integrals in four
dimensions. Nevertheless, the infrared parts of $\J_{12}$ and 
$\J_{21}$ do contain a pole at $d=4$, because the integral over $z$ in the
representation (\ref{phimnS}) only converges for $m+n>3$. We define the
renormalized infrared parts by
\bea \JSbar_{mn}\al=\al\JS_{mn}-\lambdabar\,\nu_{mn}\fs\nonumber\eea
For $m+n>2$, the full integral converges, so that
\bea \JRbar_{mn}\al=\al \JR_{mn}+\lambdabar\,\nu_{mn}\hspace{2em}
\mbox{if}\hspace{1em}m+n>2\fs\nonumber\eea

As shown in section \ref{generalization},
$\J_{mn}$, $\JS_{mn}$ and $\JR_{mn}$ may be represented as integrals over a 
derivative of $\J$, $\JS$ and $\JR$, respectively. 
The counter terms for $\JS_{mn}$ or $\JR_{mn}$ are thus given by 
a derivative of $\Jdiv$ with respect to the masses. 
In accord with the
statements made above, $\Jdiv$ is linear in $\Mp^2$ and $\Mbare^2$, so that
\bdm \Jdiv_{mn}=0
\hspace{2em} \mbox{if}\hspace{1em}m>2\hspace{0.5em} \mbox{or}\hspace{0.5em} 
n>2\fs\edm
In the case of $\JS_{12}$ (triangle graph with one meson and two nucleon
propagators), the 
formula (\ref{phiJS}) involves a single Feynman parameter:
\bea\JS_{12}\al=\al -\int_0^1\!\! dy\;
\frac{\partial\JS}{\partial \bar{B}}\left(\rule{0em}{0.7em}
(\bar{P}-\bar{q})^2,\bar{A},\bar{B}\right)\co
  \nonumber\\
 q_1\al=\al\bar{q}= 0\co\hspace{0.5em} \bar{P}=y\,P_1
 +(1-y)\,P_2\co\hspace{0.5em}  \nonumber\\
\bar{A}\al=\al \Mp^2\co\hspace{0.5em}\bar{B}=\Mbare^2-y(1-y)(P_1-P_2)^2
\fs\nonumber\eea 
The counter term is thus given by
\bea\Jdiv_{12}\al=\al -\int_0^1\!\!\frac{dy}
{\left\{\rule{0em}{0.7em}y\,P_1+(1-y)\,P_2\right\}^2}=
-\frac{1}{(P_1\!\cdot\! P_2)}\;
\frac{\arctanh\,\eta}{\eta}\co\nonumber\\
\eta\al=\al\frac{\sqrt{(P_1 \!\cdot\! P_2)^2-P_1^2P_2^2}}{(P_1\!\cdot\!
  P_2)}\fs\nonumber\eea 
As was to be expected on general grounds, the counter term does not contain any
infrared singularities. Since $\eta$ represents a term of $O(p)$, the chiral
expansion starts with
\bea \Jdiv_{12}=- \frac{1}{(P_1\!\cdot\! P_2)}+O(p^2)\fs\nonumber\eea

The renormalization of the function $\JS_{21}$ (triangle with two meson and
one nucleon propagators) may be worked out in the same manner. The
corresponding counter term is given by 
\bea\label{nu21} \Jdiv_{21}= -\Jdiv_{12}\hspace{0.2em}
\rule[-0.5em]{0.025em}{1.3em}_{\;\;P_1\rightarrow
  P\co\hspace{0.5em}P_2\rightarrow P-q}
\eea
This completes the list of renormalizations for the scalar loop integrals.

\section{Convergence of the chiral expansion,\\ explicit representations}
\label{convergence}

In section \ref{selfenergy},
we have shown that, in the case of the self energy, the chiral expansion
of the infrared part converges if the variable
$\Omega$ is in the range (\ref{range2}). We now wish to show that
the expansion actually converges throughout the low energy region.
This is most easily done on the basis of an explicit representation.

\subsubsection*{\thesection.1 Explicit representation of the self energy}
\label{con1}
At $d=4$, the integral remaining on the r.h.s.~of eq.~(\ref{Jpole}) is
elementary. In terms of the variables $\Omega$ and $\alpha$ of
eq.~(\ref{Omega}), the result for the renormalized infrared part reads
($-1\!<\!\Omega\!<\!1$) \bea \label{JSren}
\JSbar\al=\al-\frac{1}{8\pi^2}\,\frac{\alpha
  \sqrt{\rule{0em}{0.85em}1-\Omega^2}} {1+2\, \alpha\, \Omega + \alpha^2}\;
\mbox{arccos}\left(-\frac{\Omega+\alpha}{\sqrt{1+2\, \alpha\,
      \Omega +  \alpha^2}}\right)\\
\al\al -\frac{1}{16\pi^2}\,\frac{\alpha(\Omega+\alpha)} {1+2\, \alpha\, \Omega
  + \alpha^2} \left(2\,\mbox{ln}\,\alpha-1\right)\fs\nonumber\eea In accord
with the power counting of HB\ch PT, the chiral expansion of $\JSbar$ starts
at order $p$. The coefficients are nontrivial functions of $\Omega$:
\bdm\JSbar=-\frac{\alpha}{16\pi^2}\;
\left\{2\,\sqrt{1-\Omega^2}\;\mbox{arccos}\,(-\Omega)
  +\Omega\,(2\,\mbox{ln}\,\alpha-1)\right\}+O(\alpha^2)\fs\edm For the regular
part and for the full integral, the explicit expressions may be written in the
form ($-1\!<\!\Omega\!<\!1$): \bea \JRbar\al=\al
\frac{1}{8\pi^2}\,\frac{\alpha\sqrt{1-\Omega^2}}{1 + 2\,\alpha\Omega
  +\alpha^2}\,\mbox{arcsin}\left(\frac{\alpha\, \sqrt{1-\Omega^2}}
  {\sqrt{1+2\,\alpha\Omega +\alpha^2}}\right) \nonumber\\
\al\al+\,\frac{1}{16\pi^2}\,\frac{1+\alpha\,\Omega}
{1+2\,\alpha\Omega +\alpha^2}\co\nonumber\\
\Jbar\al=\al-\frac{1}{8\pi^2}\,\frac{\alpha
  \sqrt{\rule{0em}{0.85em}1-\Omega^2}} {1+2\, \alpha\, \Omega +
  \alpha^2}\; \mbox{arccos}(-\Omega )\nonumber\\
\al\al -\frac{1}{8\pi^2}\, \frac{\alpha(\Omega+\alpha)} {1+2\, \alpha\, \Omega
  + \alpha^2} \;\mbox{ln}\,\alpha +\frac{1}{16\pi^2}\fs\nonumber\eea 
In agreement with eq.~(\ref{Jhexp}), the chiral expansion of the regular part
starts at $O(p^0)$ and only contains polynomials,
\bdm\JRbar=\frac{1}{16\pi^2}\left\{1-\alpha\,\Omega+\alpha^2
\right\}+O(\alpha^3)\fs\edm 

\subsubsection*{\thesection.2 Chiral expansion of the self energy}
\label{con2}
Consider now the expansion of the function $\Jbar$ in powers of $\alpha$
at fixed $\Omega$.
The radius of convergence is
determined by the zeros of the denominator, which occur at
$\alpha=-\Omega\pm\sqrt{\Omega^2-1}$. Hence the chiral expansion of $\Jbar$
converges in the disk (\ref{disk}) -- the convergence region is the same
as the one relevant for $\JR$. In view of $\JSbar=\Jbar-\JRbar$,
the statement also holds for the expansion of the infrared part. 
This proves the claim made above.

It is essential here that we consider the chiral expansion at fixed $\Omega$.
If we instead set $P=\Mbare \,v + r$, expand in powers of $r$ and $\Mp$  and
collect terms of the same order, 
a phenomenon similar to the one 
encountered in the chiral expansion of $\gamma(t)$ occurs. 
The ordering of the double series amounts to setting $r=M\,\bar{r}$ and 
expanding in powers of $M$ at fixed $\bar{r}$. 
The parametrization implies
\bdm \Omega=v\cdot \bar{r}+\frac{M}{2 m}\,(\bar{r}^2-1)\fs\edm
In contrast to the expansion considered above, where $\Omega$ stays
put, we are now expanding this variable around the value 
$v\cdot \bar{r}$. 

The loop integral has a branch point at threshold, $P^2=(\Mbare+
\Mp)^2$. In the variable $\Omega$, this singularity occurs at $\Omega=1$. 
In the limit $M\rightarrow 0$, the branch point is mapped into the plane 
$v\cdot\bar{r}=1$. Accordingly, the radius of
convergence becomes small if $\bar{r}$ happens to be close to
this plane: The coefficients of the expansion blow up if $v\cdot \bar{r}$ 
tends to 1. In other words, the
series only converges in part of the low energy region.
This illustrates the fact that the convergence of the 
nonrelativistic expansion is a rather delicate matter,
sensitive to the details of the infrared structure.

\subsubsection*{\thesection.3 Other one loop integrals}
\label{con3}
As shown in section \ref{generalization}.2, there is a range of external
momenta where the chiral expansion converges, for all one loop graphs. The
example of the triangle graph shows, however, that the low energy region in
general contains holes, where the chiral expansion breaks down.  Even for the
self energy, we need to order the series suitably for the expansion to
converge throughout the low energy region. In that case, this can be done by
expanding at fixed $\Omega$. We do not know of a corresponding set of
variables for the general loop integral.

Our method does not rely on the chiral expansion of the loop integrals -- 
on the contrary, the problems afflicting that expansion motivated  
the present work. We have reformulated HB\ch PT in such a manner that the 
relevant infrared singularities
are summed up. 

For loop integrals with more than two vertices, the explicit 
representation becomes complicated. As is well known, all of the one loop
graphs can be expressed in terms of dilogarithms. 
A much simpler representation
may, however, be given if we resort to the approximation discussed in section
\ref{threshold}, which amounts to summing up only the {\it leading} 
infrared singularities. There, we studied the low energy properties of
the function $\gamma(t)$ -- in the above terminology, this function coincides
with $\J_{21}$, except that the two external nucleon momenta 
are put on the mass shell ($P^2=\Mbare^2$,
 $q^2=t$, $Pq=\frac{1}{2}\, t$). In that case, we considered two 
different expansions, one at fixed $t/\Mp^2$, the other at fixed 
$(t-4\Mp^2)/\Mp^4$ and then joined the two. For a corresponding 
 representation of the infrared part, we refer to the
appendix.
An alternative procedure might be to look
for a uniformizing variable replacing $t$. The breakdown of the chiral
expansion is generated by 
the fact that the form factor contains a branch point both on the first and 
on the second sheet and that the two move together if the mass of the
nucleon is sent to infinity.

As shown in appendix \ref{Loop integrals}, the sum of the 
{\it leading} infrared singularities of all one loop integrals that are
relevant for the scalar form factor and for the elastic $\pi N$ scattering
amplitude can be represented in terms of elementary functions, but we cannot
offer a general method that would lead to such a representation.
In a given
case, the issue may be settled by trial and error, for instance by 
numerically comparing the infrared part in the kinematic region of
interest with the first one or two terms in the chiral expansion thereof.
If the expansion fails, we may search for an improved approximation --
this is what we did in the case of the triangle graph. 

\section{Chiral expansion of the nucleon mass}\label{nucleon mass}
\begin{figure}[h]\centering
 \begin{tabular}{ccccc}
  \includegraphics[width=1.2in]{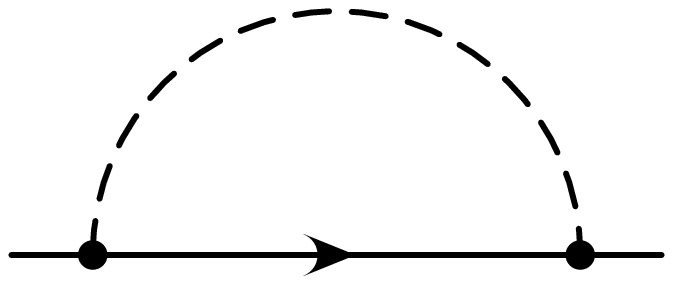} &
\hspace{0.2in} &
  \includegraphics[width=1.2in]{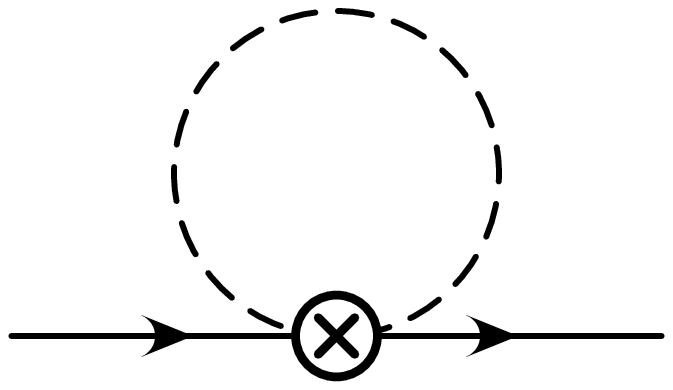} &
\hspace{0.2in} &
  \includegraphics[width=1.2in]{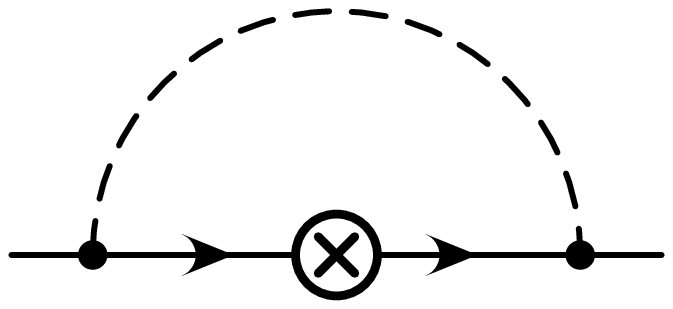}\\
a& &b& &c
 \end{tabular}
    \caption{\label{selfE2} One loop graphs contributing to the self energy of
      the nucleon. The cross denotes a vertex from ${\cal L}_N^{(2)}$.}.
\end{figure}

As a first illustration of our method, we now evaluate the physical 
mass of the nucleon to order $p^4$.
The two-point-function of the field $\psi(x)$ may be represented 
in the form
\bea i\!\!\int\!\! d^dx\,e^{iPx}\langle 0|T\,\psi(x)\,\bar{\psi}(0)|0\rangle=
\frac{1}{m+\Sigma-\Pslash}\fs\nonumber\eea
The leading contribution to $\Sigma=\Sigma(P)$ is of order $p^2$ -- 
it stems from the
term $4\,c_1 M^2\bar{\psi}\psi$ contained in ${\cal L}^{(2)}_N$. 
The one loop graphs shown in fig.~\ref{selfE2} 
are generated by ${\cal L}^{(1)}_N$, ${\cal L}^{(2)}_N$ and start
contributing at order $p^3$. Finally, there is a tree graph
contribution from 
${\cal L}^{(4)}_N$ (see eq.~(\ref{Lag4})):
\bea \Sigma=-4\, c_1 M^2 +\Sigma_a +\Sigma_b + \Sigma_c +  e_1 M^4 
+O(p^5)\fs
\nonumber\eea
The explicit expressions for the loop contributions are obtained with
the standard rules of relativistic perturbation theory:
\bea\Sigma_a\al=\al\frac{3\,g_A^2}{4\,F^2}\,(m+\Pslash )\left\{
M^2  I+(m-\Pslash)\,\Pslash I^{(1)}-\DN \right\}\co\nonumber\\ 
\Sigma_b\al=\al \frac{3M^2\Dpi}{F^2}\left\{2 c_1 -\frac{P^2}{m^2 d}\,c_2
  -c_3\right\}\co \nonumber\\
\Sigma_c\al=\al -4 c_1 M^2\, \frac{\partial\Sigma_a}{\partial m}\fs
\nonumber\eea
The mass insertion in graph c arises from the shift 
$m\rightarrow m_2=m-4 c_1 M^2$
in the nucleon propagator, generated by the term from ${\cal L}^{(2)}_N$
mentioned above. 
We could have replaced the mass in the free part of the Lagrangian by $m_2$, 
so that the graph \ref{selfE2}c would then be absent. 

The only difference to the standard evaluation of the graphs is that the
loop integrals are regularized in a different manner. In particular,
the full scalar self energy integral (fig.~\ref{selfE2}a) 
is replaced by the infrared 
part thereof, $I=I(P^2,M^2,m^2)$ -- we have discussed the properties of 
this function in detail in the preceding sections. 
The term $\DN$ denotes the scalar
nucleon propagator at the origin,
\bea \DN=\frac{1}{i}\int_I \frac{d^dk}{(2\pi)^d}\,\frac{1}
{m^2-k^2-i\epsilon}\fs\nonumber\eea
In infrared regularization, that term vanishes, because it does not 
contain any infrared singularities (see section \ref{generalization}) :
\bea \DN=0\fs\nonumber\eea
For purely mesonic loops such as the one occurring in fig.~\ref{selfE2}b, 
there is no difference between dimensional and infrared regularization:
The integral $\int_I d^dk$ coincides with the ordinary $d$-dimensional 
integral. The graph is proportional to the pion propagator at the origin, 
which we denote by $\Dpi$,
\bea \Dpi =\frac{1}{i}\int_I 
\frac{d^dk}{(2\pi)^d}\,\frac{1}
{M^2-k^2-i\epsilon}=2M^2\left(\bar{\lambda} +
\frac{1}{16\pi^2}\,\mbox{ln}\,\frac{M}{m}\right)\fs\nonumber\eea
Finally, the integral $I^{(1)}$ may be expressed in terms of $I$
(see appendix \ref{Loop integrals}) :
\bea I^{(1)}=\frac{1}{2P^2}\,\left\{(P^2-m^2 + M^2)\,\JS +\Dpi-\DN\right\}
\fs\nonumber\eea
The same relation also holds in dimensional regularization. 

The physical mass of the nucleon, which we denote by $m_N$, is determined 
by the position of the pole in the two-point-function.
In view of the factor $m-\Pslash$, the term $I^{(1)}$ does not contribute.
Evaluating the quantity $I$ with the explicit expression in
eq.~(\ref{JSren}), we obtain
\bea\label{mN} m_N\al=\al m-4\,c_1 M^2 -\frac{3\,g_A^2\, M^3}{32\pi
  F^2}+k_1\,M^4\,\mbox{ln}\,\frac{M}{m} +k_2\,M^4+O(M^5)\co\nonumber\\
k_1\al=\al-\frac{3}{32 \pi^2 F^2m}
\left(g_A^2-8\,c_1 m +c_2 m +4\,c_3 m\right)\co\\
k_2\al=\al \bar{e}_1
-\frac{3}{128\pi^2F^2m}\left(2\,g_A^2-c_2 m \right)\fs
\nonumber\eea
The formula agrees with the result of 
refs.~\cite{Steininger,Kambor Mojzis}.
The bare constants $M$, $F$, $m$, $g_A$, $c_1$, $c_2$, $c_3$, $c_4$ 
remain finite when the
regularization is removed, but $e_1$ contains a
pole at $d=4$. The quantity $\bar{e}_1$ is the corresponding renormalized 
coupling constant at scale $\mu=m$ :
\bea\label{rene1} \bar{e}_1 =e_1- \frac{3\lambdabar}{2F^2 m}
\left(g_A^2-8\,c_1 m+c_2 m+4\,c_3 m\right)\fs\eea

\section{Wave function renormalization}\label{wave function}
The wave function renormalization constant is the residue of the pole in
the two-point-function and is determined by a derivative of the self energy
with respect to the momentum,
\bea
Z^{-1}=1-\frac{\partial\Sigma}{\partial\Pslash}\hspace{0.3em}\rule[-0.9em]
{0.02em}{2.3em}_{\;P\hspace{-0.4em}\slash\,\,=\,m_N}\fs\nonumber
\eea
With the above expression for $\Sigma$,
which is valid to order $p^4$, we can extract the residue to accuracy
$p^3$. The result,
\bea Z=1-\frac{9\,M^2 g_A^2}{2 F^2}\left\{\lambdabar +\frac{1}{16\pi^2}
\left(\mbox{ln}\frac{M}{m}+\frac{1}{3}-\frac{\pi M}{2\,m}\right)\right\}+
O(M^4)\co\nonumber\eea
is in agreement with those obtained within the heavy baryon
formalism. For a detailed discussion of the latter, see 
ref.~\cite{Kambor Mojzis}.

Note that the multiplicative renormalization
$\psi^{ren}=Z^{-\frac{1}{2}} \psi$ does not render the
two-point-function finite at $d=4$. The reason is the following.
We may collect all of the correlation functions associated with
$\psi$ and $\bar{\psi}$ by adding a term of the form $\bar{\eta}\,\psi
+\bar{\psi}\,\eta$ to the effective Lagrangian, where $\eta(x)$ is an 
anticommuting
external field. Such a term, however, breaks chiral
symmetry: Under a chiral rotation, the field $\psi$ transforms with
a matrix that involves the pion field. We cannot subject 
$\eta$ to the same rotation, because this field stays put when the meson
variables are integrated out. So, off the mass shell, the correlation functions
of the effective field are regularization dependent objects.
In the context of
the effective field theory, these are without significance --
the field $\psi(x)$ merely represents a variable of integration.

Instead we could consider the field $\Psi=u\,\psi_R+u^\dagger\, \psi_L$, which
does transform with a factor that is independent of the meson field:
$\Psi_R\rightarrow V_R\,\Psi_R,$ $\Psi_L\rightarrow V_L\,\Psi_L$.
Accordingly, the counter terms needed to renormalize the quantum fluctuations
generated by the term $\bar{\eta}\,\Psi +\bar{\Psi}\,\eta$
are chirally invariant. In contrast to the correlation functions of 
$\psi$, those of $\Psi$ can unambiguously be worked out. Indeed,
these variables are relevant for the low energy analysis
of QCD operators that are formed with three quark
fields and carry the quantum numbers of the nucleon: The leading term in 
the effective field theory representation of such an operator 
is a multiple of $\Psi$.

The main point is that the correlation functions of $\psi$
do not represent physical quantities. The graphs
for say the scalar form factor or the nucleon mass
also involve nucleons propagating off the mass shell, but the result is
uniquely determined by the Lagrangian: The low energy representation of these
quantities in terms of the renormalized coupling constants 
is independent of the regularization used. The one for the
correlation functions of the effective field is unambiguous only on
the mass shell. On-shell matrix elements such as 
form factors or scattering amplitudes may be obtained by extracting
the residues of the relevant poles in suitable correlation functions --
in that connection, only the on-shell properties matter.

\section{Scalar form factor to order $p^4$}
\label{formfactor} 
In section \ref{scalar}, we considered the low energy representation of the
scalar form factor to $O(p^3)$. We now extend that representation to the next
order of the expansion (for an analogous calculation of the scalar form
factors within SU(3), based on HB\ch PT, see ref.~\cite{borasoy}). For this
purpose, we treat the quark masses $m_u,\,m_d$ as external fields and
calculate the response of the transition amplitude $\langle
N(P',s')\,\mbox{out}|N(P,s)\,\mbox{in}\rangle$ to a local change in these
fields. Within the effective theory, the transition amplitude may be worked
out by treating the term $\chi$ in the effective Lagrangian as a space-time
dependent quantity and evaluating the two-point-function $\langle
0\;\mbox{out}| T \psi(x)\,\bar{\psi}(y)|\,0\;\mbox{in}\rangle$ in the presence
of this external field: The transition amplitude is determined by the residue
of the double pole occurring in the Fourier transform of this quantity at
$P^2=m_N^2$, $P^{\prime\,2}=m_N^2$. To extract the residue, we amputate the
external nucleon legs, evaluate the remainder at $P^2=P^{\prime\,2}=m_N^2$ and
multiply the result with the wave function renormalization constant $Z$.
\subsubsection*{\thesection.1 Evaluation of the graphs}\label{sff1}
\begin{figure}[h]\centering
 \begin{tabular}{ccccc}
  \includegraphics[width=1.5in]{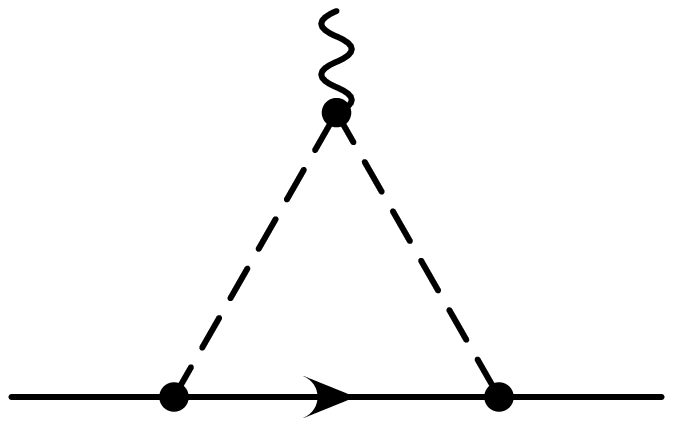} &
\hspace{0.1in} &\includegraphics[width=1.5in]{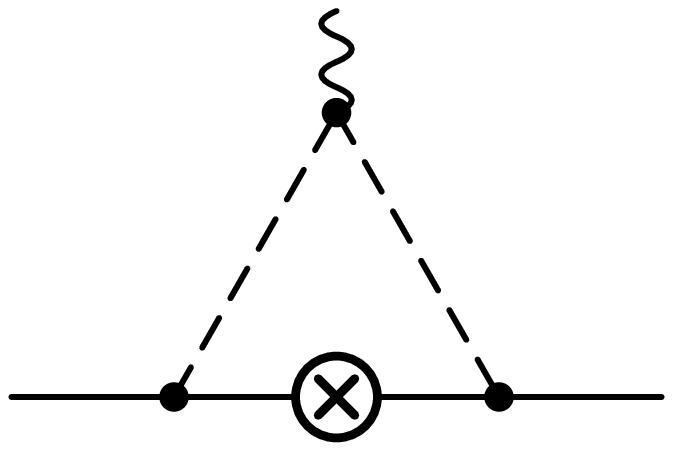}
  &\hspace{0.1in} & \includegraphics[width=1.5in]{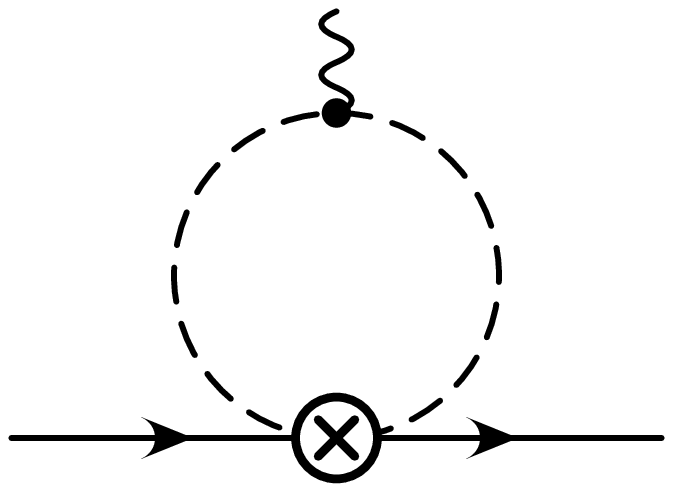} \\
a& &b && c\\ && \\ && \\
 \end{tabular}
\begin{tabular}{ccc}
  \includegraphics[width=1.5in]{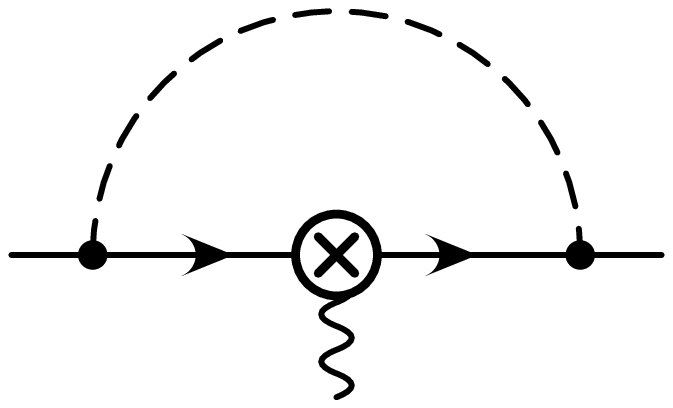}&\hspace{0.2in} &
\includegraphics[width=1.5in]{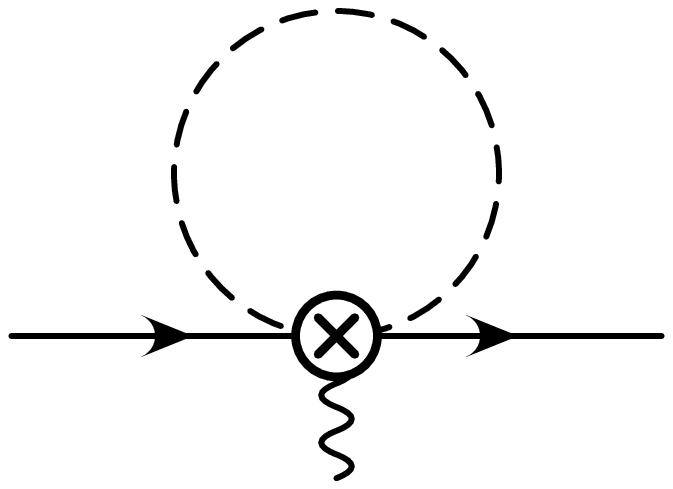}  \\
d& &e 
 \end{tabular}
    \caption{\label{cap:sigma} The loop graphs contributing to the scalar
      form factor at $O(p^4)$.}
\end{figure}

The tree graph contributions from ${\cal L}_N^{(2)}$ and 
${\cal L}_N^{(4)}$ read:
\bea \sigma_{tree}=Z\,( - 4 c_1 M^2 + 2\,e_1 M^4 + e_2 M^2 t)\fs\nonumber\eea
The one loop graphs are shown in fig.~\ref{cap:sigma}.
All of these, except the one in fig.~\ref{cap:sigma}a represent contributions
of $O(p^4)$.
When performing the perturbative calculation, it is convenient to include the 
term $4\, c_1 M^2\bar{\psi}\,\psi$ 
in the free part of the Lagrangian, 
replacing $m$ by $m_2= m - 4\, c_1 M^2$. The nucleons
occurring in the various graphs then propagate with $m_2$. Note that
the difference $m^2-P^2$ represents a quantity of order $p$. In the
propagator, the shift $m\rightarrow m_2$ thus generates a first order
correction.  The distinction between $m$, $m_2$ and $m_N$ only matters in 
the triangle diagram \ref{cap:sigma}a. In fact, graph \ref{cap:sigma}b 
represents the change occurring in this 
diagram if $m$ is replaced by $m_2$ -- this graph is absent if the mass
insertion from $c_1$ is included in the free part of the Lagrangian. 
In view of $m_N-m_2=O(p^3)$ we may also replace the factors of $m_N$ 
arising from
the matrix elements $\bar{u}'\Pslash u$ and $\bar{u}'\Pslash' u$ by $m_2$.  
The contributions from the one loop diagrams then take the form:
\bea\al\al \sigma_a+\sigma_b=-\frac{3M^2 g_A^2 }{2 F^2}\,
\{4 m^3 I_{21}^{(1)}(t)-
m\,J(t)\}\rule[-0.4em]{0em}{1.5em}_{m\rightarrow m_2}\co\nonumber\\
\al\al\sigma_c=-\frac{6\, c_1 M^4}{F^2}\, J(t)
-\frac{3\, c_2M^2}{8 F^2 m^2}\,\left\{
2\,t\,(4m^2-t)\,J^{(1)}(t) - 2\, t^2
J^{(2)}(t)+t^2 J(t)\right\}\nonumber\\
\al\al\hspace{2em} -\frac{3\,c_3 M^2}{2 F^2}\,\left\{(t-2M^2)\,J(t) + 2\,
\Delta_\pi\right\}\co\nonumber\\
\al\al\sigma_d= \frac{6\,c_1 M^2}{F^2}\,\Delta_\pi\co\nonumber\\
\al\al\sigma_e= \frac{3\, c_1 g_A^2 M^2}{F^2}\,
\left\{4 M^2 m^2 I_{12}(t)-4m^2
I^{(1)}(m^2)+\Delta_\pi\right\}\fs\nonumber\eea
The notation is specified in appendix \ref{Loop integrals}. With the
relations given there, 
the various invariants may be expressed in terms of the basic functions 
$J(t)$, $I(s)$, $I_{21}(t)$ and $I_{12}(t)$. The individual terms 
entering the combination $4 m^3 I_{21}^{(1)}(t)-m\,J(t)$ are of order
$p^0$, but the leading terms cancel: The low energy expansion of 
$\sigma_a$ only starts at $O(p^3)$, in accord with the counting of powers
in graph \ref{cap:sigma}a. Hence we may ignore the 
difference between $m_2$ and $m$ in the above expression.

The loop graphs contain divergences proportional to $M^4$ and to $M^2 t$,
respectively. The 
renormalization (\ref{rene1}) of the coupling constant $e_1$
removes the first one. The second one requires the following 
renormalization of $e_2$:
\bea
\bar{e}_2= e_2+\frac{1}{2F^2 m}\left\{3\,g_A^2+c_2 m +6\,c_3 m\right\}
\lambdaN \fs\nonumber\eea
\subsubsection*{\thesection.2 Result}\label{sff2}
We write the result for $\sigma(t)=\sigma_{tree}+\sigma_a+\sigma_b+\sigma_c
+\sigma_d + \sigma_e$ in the form
\bea \sigma(t)=\sigmaN + \tilde{\sigma}(t)\co\nonumber\eea
where $\sigmaN\equiv \sigma(0)$ represents the value of the form factor 
at the origin 
and is referred to as the $\sigma$-term. 
According to the Feynman-Hellmann theorem, this term represents
the derivative of the nucleon mass with respect to the quark masses $m_u$,
$m_d$, or, equivalently,
\bea \sigmaN=M^2\,\frac{\partial\, m_N}{\partial M^2}\fs\nonumber\eea
Indeed, the sum of the contributions from the various graphs
agrees with the derivative of the formula (\ref{mN}) for the
nucleon mass. 

For the remainder, the calculation yields: \bea\label{sigmat}
\tilde{\sigma}(t)\al=\al\frac{3\,g_A^2 M^2
  m}{4 F^2}\,\{(t-2M^2)\,\bar{I}_{21}(t)+2 M^2 \bar{I}_{21}(0)\}\\
\al\al
+ \{k_3\, M^4+k_4\,  M^2 t\}\, \bar{J}(t)+ k_5\, M^2 t+O(p^5) \co\nonumber\\
k_3\al =\al -\frac{1}{F^2}\left\{6\, c_1 - c_2 - 3\, c_3\right\}\co\nonumber\\
k_4\al =\al -\frac{1}{8 F^2 m}\left\{3\,g_A^2+2\,c_2 m + 12\, c_3 m
\right\}\co\nonumber\\
k_5\al =\al -\frac{1}{8 \pi^2}\,k_4\,\mbox{ln}\,\alpha +\frac{1}{384\pi^2F^2
  m}\left\{9\,g_A^2+2\,c_2 m + 36\, c_3 m \right\} +\bar{e}_2 \fs\nonumber
\eea 
The only difference to the result of an analogous calculation in HB\ch PT
is that the representation for the function $\bar{I}_{21}(t)$ given in
appendix \ref{Loop integrals} also covers the vicinity of $t=4 M^2$, where the
heavy baryon representation fails. The value of $\sigma(t)$ at the 
Cheng-Dashen point $t=2M^2$ was given earlier, in ref.~\cite{BKM PLB389} -- our
expression confirms this result. 

\subsubsection*{\thesection.3 Unitarity}\label{sff3}
Unitarity offers an instructive test of the momentum dependence.
Within the effective theory, the contributions from intermediate
states containing more than two pions only show up at three loop order. 
Hence the representation obtained for the form factor must obey the elastic
unitarity relation \cite{GLS}
\bea\label{unitarity}
\mbox{Im}\,\sigma(t)=\theta(t-4\,M_\pi^2)\;
\frac{3\,\sqrt{t-4\,M_\pi^2}}{2\,\sqrt{t}\;(4m_N^2-t)}
\;\sigma_\pi^\star(t)\,
f^0_+(t)+O(p^7)\co\eea
where $\sigma_\pi(t)$ is the form factor associated with the 
$\sigma$-term of the pion,  
\bea \sigma_\pi(t)=\langle\pi^0(p')|\hat{m}\,(\bar{u}\,u+\bar{d}\,d)|
\pi^0(p)\rangle\nonumber\eea and $f^0_+(t)$ is the t-channel $I=J=0$ 
partial wave amplitude of $\pi N$ scattering. 

To calculate the left hand side of the unitarity condition, we recall
that the function $\bar{I}_{21}(t)$ represents the renormalized 
infrared part of
the triangle integral $\gamma(t)$ introduced in section \ref{scalar}. 
At any finite order of the low energy expansion, the difference between the
infrared part and the full integral 
is a polynomial, so that $\mbox{Im}\,\bar{I}_{21}(t)=\mbox{Im}\,\gamma(t)$.
The explicit expression was given
in eq.~(\ref{imgamma}). For $\bar{J}(t)$, the imaginary part reads
\bea \mbox{Im}\,\bar{J}(t)=\theta(t-4M^2)\;\frac{\sqrt{t-4M^2}}
{16 \pi\sqrt{t}}\fs\nonumber\eea
The quantities on the right hand side of the unitarity condition are needed
only at tree level, where $\sigma_\pi(t)=M^2$. The corresponding
approximation for the $\pi N$ scattering amplitude is given in appendix 
\ref{scattering}. The comparison shows that the representation obtained for
the form factor indeed obeys the
unitarity condition, up to contributions that are beyond the accuracy of
a one loop calculation.

\subsubsection*{\thesection.4 Value at the Cheng-Dashen point}\label{sff4}
The low energy theorem that underlies determinations 
of the $\sigma$-term from $\pi N$  
data relates the
scattering amplitude to the scalar form factor
at the Cheng-Dashen point, where $t=2 M_\pi^2$. It is therefore
of interest to evaluate the difference 
\bea \Delta_\sigma=\sigma(2 M_\pi^2)-\sigma(0)\nonumber\eea
with the above representation of the form factor. The result is of the
form  
\bea
\Delta_\sigma\al=\al\Delta_1\,M^3 + 
\Delta_2 \,M^4\,\mbox{ln}\,\frac{M}{m}+
\Delta_3\,M^4+O(M^5)\co\nonumber\\
\Delta_1\al=\al\frac{3\, g_A^2}{64\pi F^2}\fs\nonumber\eea
The terms of order $M^4$ involve the coupling constants of ${\cal L}_N^{(2)}$:
\bea\label{Deltasigmac} \Delta_2\al=\al\frac{3\, g_A^2}{16 \pi^2 F^2 m } 
+\frac{c_2}{16\pi^2 F^2} +\frac{3\,c_3}{8\pi^2 F^2} \co \\
\Delta_3\al=\al\frac{3(2+\pi)\, g_A^2}{128\pi^2 F^2  m}-
\frac{3(4-\pi)\,c_1}{16\pi^2F^2}+\frac{(14-3\pi)\,c_2}{192\pi^2 F^2}
+\frac{3\,c_3}{16\pi^2 F^2} + 2\,\bar{e}_2 \fs\nonumber\eea
To express these in terms of observable quantities, we compare the tree 
graphs for the $\pi N$ scattering amplitude generated by ${\cal L}_N^{(1)}+
{\cal L}_N^{(2)}$ with the 
subthreshold expansion of H\"ohler and collaborators \cite{hoe} (see appendix
\ref{scattering}). Since that comparison allows us to determine the effective 
coupling constants only up to corrections of order $M$,
we denote the resulting estimates by $c_i^{(0)}$:
\bea\label{cd} c_1^{(0)}\al=\al -\frac{F^2}{4 M^2}\,
 \left\{d_{00}^+ +2 M^2\,d_{01}^+
\right\}\co\hspace{3em}
c_2^{(0)}= \frac{F^2}{2}\; d_{10}^+\\
c_3^{(0)}\al=\al - F^2\, d_{01}^+\co\hspace{10.2em}
c_4^{(0)}=\frac{1}{2 m}\left\{F^2\,b_{00}^- -\mbox{$\frac{1}{2}$}\right\} 
\fs\nonumber\eea
The above expressions for $\Delta_2$ and $\Delta_3$ then take the form 
\bea
\Delta_2\al=\al\frac{3\,g_A^2}{16\pi^2 F^2 m}+\frac{d_{10}^+}{32\pi^2}
-\frac{3\,d_{01}^+}{8\pi^2}\co\nonumber\\
\Delta_3\al=\al\frac{3(2+\pi)\, g_A^2}{128\pi^2 F^2 m}+
\frac{3(4-\pi)\,d_{00}^+}{64\pi^2  M^2 } +\frac{(14-3\pi)\,d_{10}^+}{384 \pi^2}
+\frac{3(2-\pi)\,d_{01}^+}{32\pi^2}
+ 2\bar{e}_2\fs\nonumber
\eea
The difference between
$F$, $F_\pi$ and $M$, $M_\pi$ is beyond the accuracy of the representation
(\ref{sigmat}). We replace $F$ by $F_\pi=92.4 \,\mbox{MeV}$ and use the mass
of the charged pion.  For the same reason, we may identify the 
bare coupling constant $g_A$ with the experimental value 
$g_A=1.267\pm 0.0035$ \cite{PDG}. Using the values of the
subthreshold 
coefficients quoted in appendix \ref{scattering}, we then obtain
$\Delta_1\,M^3=7.6\,\mbox{MeV}$,
$\Delta_2\,M^4\,\mbox{ln}\,\alpha=7.8\,\mbox{MeV}$ and 
$ (\Delta_3-2\, \bar{e}_2)\,M^4=
-1.4\,\mbox{MeV}$. These terms add up to
\bea \Delta_\sigma =14.0  \,\mbox{MeV}+ 2 \,M^4 \bar{e}_2\co\nonumber\eea 
to be compared with the 
result of the dispersive calculation of ref.~\cite{GLS}:
\bea\label{Deltasigmanum} \Delta_\sigma=15.2\pm 0.4\;\mbox{MeV}\fs\eea
The comparison shows that the contribution from the coupling constant
$\bar{e}_2$ is small, as it should be.
  
The above calculation resolves an old puzzle: 
The leading term in the chiral expansion of $\Delta_\sigma$ -- the one of order
$M^3$ -- accounts for only half of the result. 
The terms of order $M^4$ are numerically of the same size, because they are
enhanced by a chiral logarithm.
The underlying physics can be sorted out by noting that 
the elastic unitarity condition (\ref{unitarity}) leads to the representation
\bea\label{dIDelta} \Delta_\sigma\al=\al \int_{4M_\pi^2}^\infty dt\, n(t)
\,\sigma_\pi^\star(t)\,
f^0_+(t)\co\\
n(t)\al=\al\frac{3 M_\pi^2\,\sqrt{t-4\,M_\pi^2}}
{\pi\,t^{\frac{3}{2}}\,(t-2M_\pi^2)\,(4m_N^2-t)}\fs\nonumber\eea
The one loop calculation discussed above replaces the pion $\sigma$-term by
the first term in the chiral series, $\sigma_\pi(t)\rightarrow M_\pi^2$.
The contribution proportional to $g_A^2$ represents the value of the
dispersion integral that results if the partial wave amplitude 
is replaced by the Born approximation 
$f_{\!B+}^{\,0}(t)
=O(p)$, given in eq.~(\ref{fBorn}). The remainder is dominated by 
the
coupling constant $c_3$, which generates
a polynomial contribution\footnote{Since the corresponding contribution
to the scattering amplitude only depends on $t$, the $t$-channel 
partial wave projection is trivial and is obtained by multiplying the relevant 
term in $D^+(\nu,t)$ with the factor $(4m^2-t)/16\pi=m^2/4\pi+O(p^2)$.} 
to the partial wave amplitude: $f_+^{0}(t)_{c_3}=(2M^2-t)\,c_3\,m^2/4\pi
F^2=O(p^2)$.

\begin{figure}
\begin{center}
\psfrag{x}[lB]{\large $t$  }
\psfrag{y}[lb]{\large$n f_+^0$}
\includegraphics[width=4in]{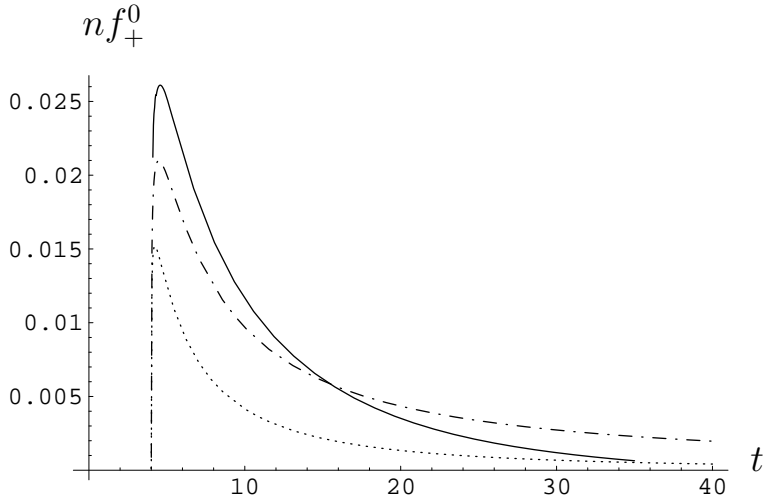}\\
\caption{\label{cap:f} The $t$-channel $I=J=0$ 
partial wave  $f_+^0(t)$. The normalization factor $n$ is chosen such that 
$\Delta_\sigma$ is the area under the curve (dotted: Born approximation,
dash-dotted: tree graphs of ${\cal L}_N^{(1)} +{\cal L}_N^{(2)}$, 
full: dispersion theory,  
all quantities in units of $M_\pi$).}
\end{center}
\end{figure}

The tree approximation for $f_+^0(t)$ is shown in
fig.~\ref{cap:f},
together with the Born term (tree graph from ${\cal L}_N^{(1)}$ alone).  To
indicate the weight in the dispersion integral (\ref{dIDelta}), we plot the
corresponding contributions to the quantity $n(t)\,f_+^0(t)$. We also depict
the result of the dispersive evaluation described in ref.~\cite{GLS}, which
includes the higher orders of the chiral series\footnote{The value quoted in
  eq.~(\ref{Deltasigmanum}) also accounts for the higher order terms in the
  form factor $\sigma_\pi(t)$, which we are ignoring here, because they start
  showing up in $\Delta_\sigma$ only at $O(p^5)$.} and is therefore complex --
the curve shown represents the quantity $n(t)\,|f^0_+(t)|$.  Qualitatively,
the picture is quite similar to the one found for the imaginary part of the
scalar form factor of the pion, $\mbox{Im}\,\sigma_\pi(t)$: The higher order
effects tend to amplify the leading order terms also in that case \cite{Gasser
  Meissner,Bijnens Colangelo Talavera}.

The figure demonstrates that, at low energies, 
the tree approximation to the scattering amplitude provides a rather
decent representation. The Born term alone, however, only dominates in the
immediate vicinity of $t=4M_\pi^2$, where it exhibits the peculiar structure
discussed in section \ref{threshold}. For $\sqrt{t}>380\, \mbox{MeV}$, the
contribution generated by the coupling constants $c_1$, $c_2$ and $c_3$ is
more important than the one proportional to $g_A^2$.

Although the straightforward expansion of $\Delta_\sigma$ in powers of the
quark masses is well-defined and can unambiguously be worked out, the first
term of that series does not yield a decent approximation. The term arises
from the infrared singularity generated by the Born approximation. On the one
hand, the contributions from the second term of the chiral expansion are
suppressed by one power of $M$, on the other, they are enhanced by a chiral
logarithm -- numerically, they are equally important.

\subsubsection*{\thesection.5 The $\sigma$-term}\label{sff5}
Finally, we turn to the value of the form factor at $t=0$.  
The first four terms in the
chiral expansion of the nucleon mass were given in eq.~(\ref{mN}). The
corresponding expansion for $\sigmaN\equiv\sigma(0)$ is of the form
\bea\label{s1} 
\sigmaN=\sigma_1 M^2  +\sigma_2M^3 +\sigma_3 M^4\,\mbox{ln}\,\frac{M}{m}
+\sigma_4 M^4  +O(M^5)\fs\eea
The coefficients follow from the Feynman-Hellmann theorem:
\bea\label{s2} \sigma_1\al=\al -4\,c_1\co\\\sigma_2\al=\al -
\frac{9\,g_A^2}{64\pi F^2}\co\nonumber\\
\sigma_3\al=\al -\frac{3}{16\pi^2F^2\, m}\,(g_A^2 -8\,c_1 m +c_2 m+
4\, c_3 m)\co\nonumber\\
\sigma_4\al=\al-\frac{3}{64\pi^2 F^2 \,m}\,
(3\,g_A^2 - 8\, c_1 m+4\,c_3 m)+2\,\bar{e}_1\fs\nonumber\eea
The tree approximation (\ref{cd}) 
for the effective coupling constants $c_1$, $c_2$,
$c_3$ suffices to work out the numerical values of $\sigma_2$, $\sigma_3$, as
well as the corresponding contribution to
$\sigma_4$. With the values of the subthreshold coefficients quoted in
appendix \ref{scattering}, we obtain $\sigma_2 M^3\!=\!-23\,
\mbox{MeV}$, $\sigma_3 M^4\,\mbox{ln}\,\alpha\!=\!-9\,\mbox{MeV}$, 
$(\sigma_4-2\,\bar{e_1})\,M^4 \!=\!0.9\,\mbox{MeV}$. 
To evaluate $\sigma_1$ in an analogous manner, we need a more accurate 
determination of
$c_1$, but we may instead take the experimental 
value $\sigmaN\!=\! 45 \,\mbox{MeV}$ from ref.~\cite{GLS}. This leads to the
estimate  
$\sigma_1 M^2+2\,\bar{e}_1 M^4\!=\! 76 \,\mbox{MeV}$, subject to an 
uncertainty of about $8\,\mbox{MeV}$.   

These numbers show that the expansion of $\sigmaN$ in powers of the quark
masses contains a large contribution from the infrared singularity
generated by the Born term, $\sigma_2 M^3$. At the next order of the expansion,
there is a logarithmic infrared singularity,
$\sigma_3 M^4\,\mbox{ln}\,\alpha $. The comparison with 
the leading term $\sigma_1 M^2$ shows that this effect is of the size typical 
for chiral logarithms. The main difference to the situation encountered in the
mesonic sector is that the expansion contains odd as well as even
powers of $p$. If the series is truncated at order $p^3$, we must expect
a less accurate representation than the one obtained in the mesonic sector at
one loop. 

The determination of the coupling constant $c_1$ to the accuracy needed here
requires a calculation of the $\pi N$ scattering amplitude to order $p^4$ and
is beyond the scope of the present paper. We can, however, study the effects
generated by the leading infrared singularities in the values of the effective
coupling constants. For this purpose, we make use of the results reported in
ref.~\cite{gss,moj,Fettes}, where the scattering amplitude is evaluated to
order $p^3$. The formulae for the coefficients of the subthreshold expansion
given in ref.~\cite{Fettes,bernard kaiser meissner} imply that the corrections
of $O(p^3)$ generate the following shifts in the values of the effective
coupling constants:
\bea\label{cp3} c_i^{(1)}\al=\al c_i^{(0)}+\delta c_i\\
\delta c_1\al=\al-\frac{59\, g_A^2 }{1536\pi F^2}\,M\co\hspace{4.5em} \delta
c_2 =\frac{5\,g_A^4+4}{64\pi F^2}\,M\co
\nonumber\\
\delta c_3\al=\al -\frac{g_A^2(48\,g_A^2 +77)}{768\pi F^2}\,M \co\hspace{2em}
\delta c_4 = \frac{g_A^2(g_A^2 + 1)}{16\pi F^2}\,M \fs\nonumber\eea As a
check, we have applied the method discussed in section \ref{generalization}
to the loop integrals that occur in the representation of
the scattering amplitude given in ref.~\cite{gss}. In the vicinity of the
point $\nu=t=0$, the chiral expansion converges for all of these (see appendix
\ref{Loop integrals}). The result confirms the formulae for the coefficients
of the subthreshold expansion quoted above and thus also corroborates the
expressions (\ref{cp3}) for the first order shifts in the effective coupling
constants. The one for $c_1$ allows us to establish the relation between the
$\sigma$-term and the subthreshold coefficients to accuracy $p^3$:
\bea\sigmaN=F_\pi^2\,(d_{00}^++ 2 M_\pi^2 d_{01}^+) + \frac{5\,g_A^2
  M_\pi^3}{384\pi F_\pi^2} +O(M_\pi^4)\co\nonumber\eea to be compared with the
prediction at tree level, $\sigmaN=F_\pi^2\,(d_{00}^++ 2 M_\pi^2 d_{01}^+)$,
where the third term is missing. Numerically, the shift amounts to only
$2.1\,\mbox{MeV}$: In the relation between the observables $\sigmaN$,
$d_{00}^+$ and $d_{01}^+$, the effects generated by the leading infrared
singularities are an order of magnitude smaller than those seen in the chiral
expansion of $\sigmaN$. This demonstrates that the large infrared singular
contribution that occurs in the chiral expansion of the $\sigma$-term at
$O(M^3)$ also appears in the $\pi N$ scattering amplitude -- the relevant
combination of subthreshold coefficients picks up nearly the same
contribution.  Note that the magnitude of the $\sigma$-term as such is not at
issue here. For the values of the coefficients $d_{00}^+$, $d_{01}^+$ given in
ref.~\cite{hoe}, the tree level and one loop results for $\sigmaN$ are $50$
and $52\,\mbox{MeV}$, respectively. We could have used a somewhat different
input, for instance one for which the outcome for the $\sigma$-term is 45 MeV
-- the difference between the tree and one loop approximations would be the
same.

\section{Discussion}\label{discussion}
\subsubsection*{\thesection.1 Reordering of the perturbation series}
\label{disc1}
The fact that the bulk of the infrared singular contributions occurring in the
chiral expansion of the nucleon
mass and of the $\sigma$-term merely amounts to a shift of the bare parameters 
was noted long ago \cite{Quark masses}. 
In the preceding section, we listed the
change in the effective coupling constants $c_1,c_2,c_3,c_4$
generated by the infrared singularities at $O(p^3)$ and pointed out
that, in the relations between the observable quantities considered, 
these singularities nearly cancel. 
We expect a similar, albeit less dramatic reduction
also to occur in other quantities of physical interest. Some of
the fluctuations seen in the published results for the threshold parameters of
$\pi N$ scattering, for instance, merely reflect the fact that the
expansion of the various observables in powers of the quark masses contain
large contributions from infrared singularities. 

A more significant comparison
of the results obtained at tree level and at one loop results
if one considers the predictions of the effective theory: Instead of comparing
the contributions arising in the chiral expansion of a given observable
at various orders of the expansion, one may compare the relations 
between observable quantities which follow at
tree level with those obtained at one loop. 
The effective 
coupling constants entering the one loop result are different from those
relevant at tree level. The tree level values
are the best choice for the 
representation of the scattering amplitude to $O(p^2)$, 
those in eq.~(\ref{cp3}) are 
relevant for the improved representation of this amplitude 
that accounts for the terms of $O(p^3)$. 

The elimination of the 
coupling constants in favour of physical quantities
amounts to a reordering of the perturbation series. 
In more explicit terms, the reordering 
we are advocating here is the following (we restrict ourselves to a
discussion to one loop). The perturbation series relies on a decomposition of
the effective Lagrangian into a leading term and a perturbation,
\bea {\cal L}_N\al=\al{\cal L}_0 +{\cal L}_1\fs\nonumber\eea
The standard ordering results if ${\cal L}_0$ is 
identified with the first
two terms of the derivative expansion. We argue that the decomposition
\bea{\cal L}_0\al=\al{\cal L}^{(1)}_N+{\cal
  L}^{(2)}_N\,
\rule[-0.6em]{0.03em}{1.5em}_{\;c_i\rightarrow c_i^{(0)}}\nonumber\\
{\cal L}_1 \al=\al{\cal L}^{(3)}_N
+{\cal L}^{(4)}_N+{\cal L}^{(2)}_N\,
\rule[-0.6em]{0.03em}{1.5em}_{\;c_i\rightarrow \delta c_i}\co\nonumber\eea
with $\delta c_i=c_i-c_i^{(0)}$, is a better choice, because it
reduces the magnitude of the perturbation
${\cal L}_1$. Like for the standard bookkeeping, an evaluation of the 
various observables 
to $O(p^2)$ requires the
calculation of the tree graphs belonging to ${\cal L}_0$; for a
representation to $O(p^4)$, we need to add the tree graphs of ${\cal L}_1$ as
well as the one loop graphs of ${\cal L}_0$. 

Two modifications occur when the
perturbation series is extended from $O(p^2)$ to $O(p^4)$: new graphs
and the change $\delta c_i$ in the values of the coupling constants
(the explicit expressions for the latter in eq.~(\ref{cp3}) account for
the shifts of order $\delta c_i=O(M)$, but do not include those of 
order $M^2$). As discussed above, 
the two effects partly cancel. In the above bookkeeping, the change $\delta
c_i$ is booked among the corrections, on the same footing
as the contributions from the loop graphs, which both affect the
observables and are responsible for the shift in the coupling constants. 
The cancellations between the two types of contributions thus 
tend to reduce the magnitude of the perturbations generated by ${\cal L}_1$.

At first sight, the claim that the values of the coupling constants depend
on the order at which the perturbation series is considered, may appear to
contradict the fact that these constants represent perfectly well-defined
quantities that determine the chiral expansion coefficients  
of the various observables. As is well known, the results found
for these coefficients on the basis of a calculation to some given order
of chiral perturbation theory represent low energy theorems that
remain strictly valid if the expansion is carried to higher orders.
We do not put this into question, but merely emphasize that the 
result obtained for the values of the
coupling constants does depend on the order to 
which the perturbation series is worked out. 

To illustrate the need for a distinction between the coupling constants as
such and their values at a given order of the perturbation
series, we again consider the 
quantity $\Delta_\sigma$ and focus on
the chiral logarithm contained therein, $\Delta_2 M^4\,\mbox{ln}\,\alpha$. 
The term arises from the infrared singularity occurring at the lower
end of the dispersion integral (\ref{dIDelta}). 
According to eq.~(\ref{Deltasigmac}), the coefficient $\Delta_2$ contains a
piece that is proportional to the coupling constants $c_2$ and $c_3$. It
arises from the contribution to the partial wave $f^0_+(t)$ that
is generated by the tree graphs of ${\cal L}^{(2)}_N$. Hence the 
coupling constants relevant for an evaluation of $\Delta_2$ are those
pertaining to the tree level representation of the scattering amplitude.
If we were to calculate $\Delta_2$ with the improved values
of the coupling constants that follow from the relations
(\ref{cp3}), we would
in effect be using a tree level representation for $f^0_+(t)$ with the
wrong coupling constants. The point here is that the same one loop graphs that
give rise to a change in the values of $c_2$, $c_3$ also modify 
the partial wave amplitude. Since a substantial fraction of the infrared
singularities occurring at $O(p^3)$ affects the term $\Delta_\sigma$ and the
scattering amplitude in the same way, it does not make sense to account for 
one of the changes without accounting for the other.

In a certain sense, the odd and
even powers of the chiral series lead a life of their own.
The quantity $\Delta_\sigma$ illustrates the fact that the leading
terms in both of the subseries need to be investigated in order to arrive at
a significant result.
To our knowledge, the available one loop results for the $\pi N$
scattering amplitude account for the expansion only to $O(p^3)$.
It will be very interesting to compare the predictions that follow 
from the representation 
of the amplitude to $O(p^4)$   
with those obtained from the same experimental input
at tree level.

The problem also occurs in the mesonic sector. The infrared singularities
are weaker there, because the expansion only involves even
powers of $p$. At the
precision reached with a two loop calculation,
however, the need for a distinction between the coupling constants as such
and the values found at a given order of the perturbation series 
manifests itself quite clearly. As discussed in 
refs.~\cite{Bijnens et al,Stern et al}, for instance, inconsistencies may arise
if the two loop representation for the $\pi\pi$ scattering amplitude
is evaluated with the values of the coupling constants $\ell_1$, $\ell_2$
obtained on the basis of a one loop calculation from $K_{e_4}$ decay.

\subsubsection*{\thesection.2 The role of the $\Delta(1232)$}
\label{disc2}
The occurrence of a chiral logarithm explains why an evaluation of 
$\Delta_\sigma$ to order $M^3$
does not yield a decent approximation for this quantity. The coefficient of
the logarithm involves the combination $c_2 +6\,c_3$ of effective
coupling constants, which is dominated by $c_3$. It is understood why the 
value of this coupling constant obtained from low energy $\pi N$ phenomenology
is large: This constant receives an important contribution from the 
singularities
generated by the $\Delta(1232)$. In fact, it was
noted in ref.~\cite{BKMSigma} that a calculation of the scalar form
factor that explicitly includes the $\Delta$ degrees of freedom yields
a result for $\Delta_\sigma$ that is consistent with the one obtained from
dispersion theory.

The role of this state for the low energy structure in the baryonic sector is
discussed in detail in the literature
\cite{Tang,hoe,BKMSigma,Manohar,Hemmert}. For a recent review, in particular 
also of the small scale
expansion that allows one to analyze the extension of the effective theory in
a controlled manner, we refer to \cite{Kambor}.  The extended theory is
compared with the framework used in the preceding sections in appendix
\ref{delta}, where we also estimate the contributions to the effective
coupling constants that are generated by the $\Delta$.

For the quantities analyzed in the present paper,
it is not essential whether the $\Delta$ is incorporated 
as a dynamical variable in the effective Lagrangian or whether the
effects generated
by this state are accounted for only indirectly, 
through the values of the effective coupling constants:
In the domain studied here, the graphs describing the exchange of a $\Delta$
are adequately described by those terms of the chiral
expansion that occur up to and including $O(p^4)$. 

In the Mandelstam plane, the point around which the
chiral expansion is performed corresponds to $s_0=m^2 + M^2$, 
$t=0$. The expansion of the 
resonance denominators is controlled by the ratio 
$x=(s-s_0)^2/(m_\Delta^2-s_0)^2$. For small values of $s- s_0$ and $t$, in
particular near the Adler zero and for quantities like $\sigma(0)$
or $\sigma(2M^2)$, the 
expansion rapidly converges. At 
the threshold, where $x=4 M^2 m^2/(m_\Delta^2-m^2-M^2)^2\simeq 0.18$, the 
expansion of the resonance
denominators is still under good control. For  
higher energies, however, the singularities generated by the
$\Delta$ must explicitly 
be accounted for to arrive at a decent representation of the scattering
amplitude. 

In the mesonic sector, the $\rho$ plays an analogous role. Remarkably, as far
as the expansion of the resonance denominators is 
concerned, the convergence radius is roughly the same: 
$m_\Delta^2-m^2-M^2\simeq m_\rho^2-2M^2$ (the left hand side is even 
a little larger). The two Mandelstam triangles, 
however, are of very different size: The one relevant for
$\pi \pi$ scattering is smaller by the factor $M/m$, so that
the threshold is much closer to the Adler zero.
At the threshold, the parameter that controls the expansion of the term
$1/(m_\rho^2-s)$ 
is very small indeed: $x=4 M^4/(m_\rho^2-2M^2)^2\simeq 0.005$. 
In the mesonic
sector, an effective theory that does not explicitly account for 
this singularity yields meaningful results even well above threshold. 
  
\subsubsection*{\thesection.3 Comparison with the static model}\label{disc3}
It is instructive to compare the framework discussed above with
the earliest version of an effective field theory for the baryons, 
the static model. This model is characterized by the Hamiltonian 
\cite{Henley Thirring}:
\bea H=\int \!d^3\!x \,\frac{1}{2}\left\{\dot{\bphi}^{\,2}+
{\nabla \bphi}^{\,2}+{M_\pi^2\,\bphi}^{\,2}\right\}+\int\!\!d^3\!x\,
\frac{g_A}{2 F_\pi}\,\rho(\vec{x})\,\psi^\dagger
\,(\mbox{\boldmath$\sigma$}\cdot \nabla)\,
(\bphi\cdot\mbox{\boldmath$\tau$})\,\psi\fs\nonumber\eea 
The nucleon is kept fixed at the origin,
$\psi(x)\rightarrow\psi(t)$. The corresponding four-dimensional 
space of states is spanned by vectors that differ
in the spin and isospin quantum numbers -- the operator 
$\psi^\dagger \ldots \psi$ generates
transitions between these. The result for the self energy of the nucleon reads
\bea E=-\frac{3\,g_A^2}{32\pi F_\pi^2}
\int\!\! d^3\!x d^3\!y \,\,\nabla\rho(\vec{x})\cdot\nabla\rho(\vec{y})\;
\frac{e^{-M_\pi|\vec{x}-\vec{y}\,|}}{|\vec{x}-\vec{y}\,|}\fs\nonumber\eea
The function $\rho(\vec{x})$ is normalized to $\int\! d^3\!x\,
 \rho(\vec{x})=1$ --
it describes the structure of a
nucleon that is stripped of its meson cloud.
 
In the relativistic formulation of the effective theory, 
the nucleon is represented
as a point particle, $\rho(\vec{x})\rightarrow\delta(\vec{x})$,
and the leading contributions to the self energy of the 
nucleon arise from one loop graphs. To compare the expressions for the
self energy,
we note that in the static model, the expansion of the exponential
yields the series
$1-M_\pi r+\frac{1}{2}M_\pi^2 r^2-\frac{1}{6}M_\pi^3 r^3+\ldots$ The second
term vanishes upon integration, so that the series takes the form
$E=E_1+E_2 M_\pi^2 + E_3 M_\pi^3 +\ldots$ In our formulation of B\ch PT,
only the infrared part of
the loop integrals is retained, the remainder being absorbed in the effective 
coupling constants. In that language, the term $E_1$ is included in 
the bare mass $m$, while $E_2$ represents a contribution to 
the effective coupling constant $c_1$. The coefficient $E_3$ is
independent of the shape of $\rho(\vec{x})$ and the value 
$E_3=-3\,g_A^2/32\pi F_\pi^2$ indeed 
reproduces the coefficient occurring in the
chiral expansion (\ref{mN}) of the nucleon mass.
At the next order of the expansion, however, 
the static model fails: The self energy $E$ does not contain a chiral 
logarithm. In fact, in the local limit 
$\rho(\vec{x})\rightarrow\delta(\vec{x})$, the expansion
terminates at $O(M_\pi^3)$. The
deficiency arises because the model only accounts for the leading 
term ${\cal L}_N^{(1)}$ in the derivative expansion of the effective 
Lagrangian -- the formula (\ref{s2}) shows that, in the static limit 
($m\rightarrow \infty$), the nucleon mass does
not contain a chiral logarithm if the coupling constants $c_1, c_2, c_3\in
{\cal L}_N^{(2)}$ are turned off.

Quite a few other phenomena ($\pi N$ scattering, magnetic moments and
electromagnetic form factors, pion photoproduction, 
Compton scattering, nuclear forces, for instance)
were analyzed in detail within the static model, which provides a simple
intuitive picture for the basic low energy features (for an excellent overview,
we refer to the book of Henley and Thirring \cite{Henley Thirring}). 
A comparison of the
results obtained with the modern version of the effective theory with those
found in this model would be most instructive.

\subsubsection*{\thesection.4 Momentum space cutoff}\label{disc4}
The formulation of the effective theory in terms of 
pointlike nucleons has the advantage of being model independent. It
systematically accounts for all contributions
arising to a given order in the chiral expansion of the various
observables. The price to pay is that the loop integrals 
cannot be interpreted directly in
physical terms. The machinery does incorporate the finite extension of the 
nucleon, but only indirectly, through the 
effective coupling constants. A more intuitive picture of the meson
cloud that does explicitly account for the
finite size of the nucleon was proposed in ref.~\cite{Quark masses}. In that
framework, the pointlike
vertices are replaced by form factors, which in effect cut the virtual meson
momenta off. The proposal is based on the static model, 
where the $\pi N$ interaction is also equipped with a form factor,
given by the Fourier transform of the function $\rho(\vec{x})$. 
The motivation for introducing such a cutoff was that
the matrix elements of the perturbations encountered in the extension of
B\ch PT from SU(2) to SU(3) are large. 
It does not make much sense to truncate 
the straightforward expansion of the baryon masses 
in powers of $m_u,m_d$ and $m_s$ at order $p^3$.
The net effect of a cutoff is qualitatively similar to the one resulting from
the reordering discussed above: It reduces the magnitude of those terms that
are treated as perturbations. 

The proposal is taken up in the recent literature 
\cite{Donoghue Borasoy}. In particular, the quantity $\Delta_\sigma$ is
studied within that framework in ref.~\cite{Borasoy}. In the language of the
dispersion relation (\ref{dIDelta}), the calculation reported there
essentially amounts to replacing the partial wave amplitude $f_+^0(t)$ by 
the Born term (dotted line in fig.~\ref{cap:f}) and cutting the dispersion 
integral off at moderate values of $t$. Such a
calculation evidently yields a 
much smaller value for $\Delta_\sigma$ than the one that follows, either from
dispersion theory or from the evaluation of the
chiral perturbation series to order $p^4$.
The analysis of section \ref{formfactor}.4 shows that the contributions of 
order $p^3$ and $p^4$ are of a different origin. The 
terms of order $p^4$ are not properly accounted for by cutting 
off the virtual momenta in the one loop graphs of order $p^3$.

More generally, the problem with the approach proposed in 
ref.~\cite{Quark masses} is that it is model dependent. Moreover, 
introducing a cutoff in
general ruins the Ward identities of chiral symmetry. The model independent
method we are proposing does preserve the Ward identities as well as the
infrared structure. The reordering of the perturbation series also leads to a
more rapidly convergent expansion.  It yet remains to be seen, however,
whether this method will provide a coherent understanding of the mass spectrum
of the baryon octet.

\section{Conclusion}\label{conclusion}
\rule{1.5em}{0em}1. We have shown that B\ch PT can be formulated in 
such a manner that both
Lorentz invariance and chiral power counting are preserved at every stage of 
the calculation. The method relies on the fact that -- for noninteger values 
of the dimension -- the infrared singular parts of the loop graphs can 
unambiguously 
be separated from the remainder: They involve fractional powers of
the chiral expansion parameters, while the remainder admits an ordinary Taylor
series. The two parts are chirally invariant by themselves 
in the sense that they separately obey the Ward identities
of chiral symmetry. This allows us to retain only the infrared singular 
parts of the loop integrals, absorbing the remainder in the coupling 
constants of the effective Lagrangian.

2. The calculations required by our method are nearly identical to those
relevant within the relativistic formulation of the effective theory
given in ref.~\cite{gss}. The Lagrangian and the Feynman graphs are the same.
The only difference is that the loop 
integrals are replaced by the
corresponding infrared parts. At one loop order, this is achieved by  
extending one of the Feynman parameter integrations
from the interval $0<z<1$ to the interval $0<z<\infty$. 

3. We have shown that, in the sector with baryon number 1, the chiral
expansion breaks down in certain regions of phase space. The nonrelativistic
expansion that underlies HB\ch PT inherits this problem and also leads to
technical difficulties of its own, related to the fact that the
nonrelativistic expansion of infrared singularities is a rather subtle matter.
A coherent heavy baryon representation only results if the insertions required
by relativistic kinematics are summed up to all orders. The problem arises
from the interchange of the loop integration with the nonrelativistic
expansion, which is not always legitimate. The method we are proposing avoids
these difficulties ab initio, because it does not rely on a nonrelativistic
expansion of the loop integrals.
 
4. One of the advantages of chiral perturbation theory is that, at the
algebraic level, the results are unambiguous. This implies that
the chiral expansion of the representations obtained with the method proposed
here must agree with the results of HB\ch PT, order by order -- even in those
cases where this expansion does not converge.  

5. As our method avoids the nonrelativistic expansion
of the vertices, the number of graphs to be
evaluated is smaller than in HB\ch PT. The price to pay is that
the simplifications offered by nonrelativistic kinematics
cannot be made use of. As an example, we mention processes
involving external photons, such as pion photoproduction. Bernard, Kaiser,
Kambor and Meissner \cite{mei} have shown that, in the heavy baryon approach, 
many of the one loop graphs occurring in that context can be dropped: 
In the nonrelativistic limit and
in the Coulomb gauge, vertices 
involving the coupling of photons to a nucleon line are suppressed. 
Since our method leads to the same
power counting rules as the nonrelativistic expansion,
we expect these simplifications to also apply within our framework, 
but a relativistic formulation of the corresponding selection rules
yet remains to be given.

6. In the baryonic sector,
the expansion involves odd as well as even
powers of momenta. Most of the available calculations 
only account for the terms arising up to and including $O(p^3)$. 
The corresponding representation
for the observables of physical interest is inherently less accurate than the
one obtained at one loop in the mesonic sector, which holds up to and
including $O(p^4)$. A full one loop calculation of the $\pi N$ scattering
amplitude yet remains to be carried out. 

7. We have illustrated our method 
with an evaluation of the $\sigma$-term and of
the corresponding form factor to order $p^4$. The form
factor can be expressed in terms of elementary functions throughout the low
energy region. In the domain where the standard
chiral expansion in powers of momenta and quark masses is convergent, the
result agrees with the one obtained on the basis of HB\ch PT. 
The representation constructed in the present paper,
however, also holds in the vicinity of $t=4\,M_\pi^2$, where 
the heavy baryon chiral perturbation series diverges.

8. The example of the form factor shows that the expansion
of some of the observables in powers of the quark masses contains a large
contribution from the infrared singularities generated by the Born term,
proportional to $g_A^2 M_\pi^3$. We have shown, however, that
the bulk of this contribution is common to the $\sigma$-term and to the $\pi N$
scattering amplitude, so that it drops out when considering the predictions of
the theory, i.e.~the relations between the observables. 

9. Although the straightforward expansion of the various observables in
powers of momenta and quark masses is perfectly meaningful, it does not yield
a suitable ordering of the perturbation series. We have proposed a model 
independent reordering of this series, which can also be performed in 
HB\ch PT.  
At least in the two cases considered 
in the present paper ($\sigma$-term and momentum dependence of the 
scalar form factor), 
the operation strongly reduces the magnitude of the corrections.
To see whether or not 
this is a peculiarity of the observables considered here, it would be
of interest to perform the operation with the 
known representation of the $\pi N$ scattering amplitude to
$O(p^3)$ \cite{gss,moj,Fettes,Tang}. 

10. We have discussed the effects generated by the $\Delta(1232)$ in some 
detail. This state plays a role similar to the one 
of the $\rho$ in the mesonic sector: In the Mandelstam plane, the 
singularities associated with the $\Delta$ and the $\rho$ 
occur at about the same distance from the point around which the chiral 
expansion is performed. As far as the behaviour of the scattering amplitude
in the vicinity of the Adler zero is concerned, or for quantities such as
$\sigma(0)$ or $\sigma(2 M^2)$, the singularities generated by the $\Delta$
are adequately described by their contributions to the effective coupling
constants. 

11. The physical region of $\pi N$ scattering is further away from the 
Adler zero than the physical region of $\pi \pi$ scattering, 
by one power of $m/M$. Although, at the threshold, the singularities generated
by the $\Delta$ may still be replaced by their contributions to the 
effective coupling constants,
we must expect the perturbation series for the threshold 
parameters to converge less rapidly for $\pi N$ scattering than for $\pi\pi$
scattering.  

12. In the experimentally accessible region, the $\Delta$ does play a 
prominent role. The work reported in 
ref.~\cite{Leisi} indicates that, even in the vicinity of the resonance, 
a decent description of the observed behaviour of the scattering amplitude 
may be obtained by supplementing the tree graphs of a 
simple resonance model with unitarity corrections. 
It yet remains to be seen whether a systematic
analysis of the extended effective theory that includes the $\Delta$ among the
dynamical variables and incorporates chiral symmetry ab initio
will allow us to establish direct contact with the wealth of experimental
data. 

13. The problems discussed
 in the present paper
are not peculiar to the nucleons, but occur whenever the effective theory
contains degrees of freedom that remain massive in the chiral limit.
An application of our method to $\pi K$ scattering within SU(2)$\times$SU(2)
is described in ref.~\cite{Roessl} -- in that 
framework, the mass of 
the kaon sets the heavy scale. 

14.  As is well known, the 
straightforward 
expansion of the masses and current matrix elements of the
baryon octet \cite{borasoy}
and the vector mesons \cite{Bijnens rho} in powers of $m_u$, $m_d$ 
and $m_s$ contains 
large contributions
from infrared singularities (in particular terms proportional to $M_K^3$). 
It would be of considerable interest to apply
the reordering of the chiral perturbation series proposed in section
\ref{discussion}.1 to these quantities: The procedure should lead to a more
rapidly convergent expansion, so that the first few terms may then provide
a meaningful determination of the ratio $(m_d-m_u)/(m_s-\hat{m})$, 
for instance.

\vspace{2em}
\noindent{\Large\bf Acknowledgement}

\vspace{0.5em}\noindent
 We thank J\"urg Gasser for stimulating remarks and
cooperation. Also, we acknowledge
useful comments from Hans Bijnens, Gerhard Ecker, Alex Gall, 
Joachim Kambor, Ulf Meissner, 
Martin Moj\v zi\v s and Sven Steininger.
One of us (H.L.) thanks the Erwin Schr\"odinger Institute in Vienna 
for support -- part of the work was carried out during
a stay there.
   
\appendix
\renewcommand{\theequation}{\thesection.\arabic{equation}}
\setcounter{equation}{0}
\section{Low energy representation for the triangle graph}\label{A}
The function $\gamma(t)$ may be expressed in
terms of the imaginary part by means of 
the dispersion relation \cite{gss}
\bea
\gamma(t)=\gamma(0)+\frac{t}{\pi}\int_{4 \Mp^2}^{\infty}
\frac{dt'}{t'(t'-t)}\im\gamma(t')\fs
\nonumber\eea
For $t=0$, the integral (\ref{gammaloop}) can be done explicitly. 
Expanding the result in powers
of $\alpha = \Mp/\Mbare$, the subtraction constant becomes
\bea
\gamma(0) =\frac{1}{32\pi \Mbare \Mp}\left\{1+ \frac{1}{\pi}
\,\alpha \,(2\,\mbox{ln}\alpha-1) +O(\alpha^2)\right\}\fs
\nonumber\eea
For a representation of $\gamma(t)$ up to and including first nonleading
order, we need the imaginary part only to this accuracy. Inserting the
representation (\ref{repim}) and performing the integral, we
obtain
\begin{eqnarray}\al\al\hspace{-0.8em}\gamma(t)= \ghb(t) +\gth(t)+O(p)
\co\nonumber\\
\al\al\hspace{-0.8em} \ghb(t)=\frac{1}{32\pi \Mbare \Mp}
\left\{\frac{1}{\sqrt{\tau}}\;
\mbox{ln}\,\frac{2\! +\!\sqrt{\tau}}{2\! -\!\sqrt{\tau}} +
\frac{2\,\alpha\, (2\! -\!\tau)}{\pi\sqrt{\tau\,(4\! -\!\tau)}}\;\arcsin
\frac{\sqrt{\tau}}{2}+\frac{2}{\pi}
\,\alpha \,(\mbox{ln}\alpha\!-\!1)\right\},\nonumber\\
\al\al\hspace{-0.8em}\gth(t)=\frac{1}{32\pi \Mbare \Mp}
\left\{\frac{\alpha}
{\sqrt{4 \!-\!\tau}}-\mbox{ln}\left(\!1+\frac{\alpha}
{\sqrt{4\!-\!\tau}}\right)\right\}\co\nonumber \end{eqnarray}
with $\tau\equiv t/M^2$.
We have included the subtraction constant in $\ghb(t)$, so that 
not only the imaginary part of the term $\gth(t)$, but also its real
part is significant only in the immediate vicinity of
threshold. At leading order of the chiral expansion, 
the expression for $\ghb(t)$ reduces to the known result of
HB$\ch$PT \cite{mei2}. The
above formulae (i) extend that result to first nonleading order and (ii)
account for the sum of internal line insertions that determine the behaviour
in the vicinity of threshold.

In the notation used in the present paper, the function $\gamma(t)$
represents the scalar integral $\J_{21}$ on the mass shell ($q^2=t$, 
$P^2=m^2$, $Pq=\frac{1}{2}\,t$).
The above approximate
representation for this function is valid throughout the low energy region,
but is accurate only modulo contributions of $O(p)$.
For the calculation of the form factor (section
\ref{formfactor}) we need a corresponding expression for the infrared part.
For that purpose, it suffices to work out
the regular part $\JR_{21}$ to the desired accuracy. 
As discussed in section \ref{generalization}, the integral may be represented
as ($q_1=0$, $q_2=q$, 
$P_1=P$) 
\bea \JR_{21}\al =\al \frac{\Gamma(3-
\mbox{$\frac{d}{2}$})}{ (4\pi)^{\frac{d}{2}}}
\int_0^1\!\! dx  \!\int_1^\infty\!\!\!dz\, (1-z) \;
C^{\,\frac{d}{2}-3}
\co\nonumber\\
C\al =\al (1-z)\,\left\{M^2-x(1-x)\,q^2\right\} +z\,m^2 - z\,(1-z)\,
\left\{P-(1-x)\,q\right\}^2
\fs\nonumber 
\eea
The chiral expansions of the regular parts are ordinary Taylor series that
converge throughout the low energy region. Since the series only starts at
$O(p^0)$, we can truncate it at leading order, i.e.~evaluate
the above integral for $M=q=0$, $P^2=m^2$. This gives
\bea \JR_{21}= m^{d-6}\frac{2\,\Gamma(2-\mbox{$\frac{d}{2}$})}
{ (4\pi)^{\frac{d}{2}}(5-d)}+O(p)\fs\nonumber\eea
The corresponding counter term was worked out in section
\ref{renormalization}. At leading order, eq.~(\ref{nu21}) yields
$\Jdiv_{21}=m^{-2}$. Indeed the term $\lambda\,\Jdiv_{21}$ removes the
pole of the $\Gamma$-function at $d=4$. 
Renormalizing at scale $\mu=m$, we obtain
\bea \JRbar_{21}=\JR_{21}+\lambdabar\,\Jdiv_{21}=-\frac{3}{32\pi^2
  m^2}+O(p)\fs\nonumber
\eea
A corresponding representation for the renormalized infrared part 
$\bar{I}_{21}$ is obtained by
subtracting this term from the formula for $\gamma(t)$ given
above (for an explicit expression, see appendix \ref{Loop integrals}).
 
\section{Infrared parts of some loop integrals}\label{Loop integrals} 
\setcounter{equation}{0}
{\bf Notation}
\bea \lambdabar\al=\al\frac{m^{d-4}}{(4\pi)^2}\left\{\frac{1}{d-4}-\frac{1}{2}
\left(\rule{0em}{1em}\,\mbox{ln}\, 4\pi
  +\Gamma'(1)+1\right)\right\}\nonumber\\
\alpha\al=\al\frac{M}{m}\nonumber\eea
{\bf 1 meson:\; $\Dpi=I_{10}$}
\bea \Dpi =\frac{1}{i}\int_I 
\frac{d^dk}{(2\pi)^d}\,\frac{1}
{M^2-k^2}= 2M^2\left(\bar{\lambda} +
\frac{1}{16\pi^2}\,\mbox{ln}\,\alpha\right)
\nonumber\eea
{\bf 1 nucleon:\; $\DN=I_{01}$}
\bea \DN =\frac{1}{i}\int_I 
\frac{d^dk}{(2\pi)^d}\,\frac{1}
{m^2-k^2}=0\nonumber\eea
{\bf 2 mesons:\; $J=I_{20}$}
\vspace*{-0.5em} 
\bea \al\al \{J\,,\,J^\mu\,,\,J^{\mu\nu}\}=
\frac{1}{i}\int_{I}\frac{d^dk}{(2\pi)^d}\,
\frac{\{1\,,\,k^\mu\,,\,k^\mu k^\nu\}}
{(M^2-k^2)(M^2-(k-q)^2)}\rule{4em}{0em}\nonumber\\
\al\al t=q^2\co\nonumber\\
 \al\al J(t)=\bar{J}(t)- 2\,\bar{\lambda}-\frac{1}{16\pi^2}\,(
2\,\mbox{ln}\,\alpha + 1)\nonumber\\
\al\al \bar{J}(t)\equiv J(t)-J(0)=\frac{1}{8 \pi^2}
\left\{1-\sqrt{\frac{4M^2-t}{t}}\,
\arcsin\,\frac{\sqrt{t}}{2M}\right\}\nonumber\\
\al\al J^\mu=\mbox{$\frac{1}{2}$}\,q^\mu\,J(t)\rule{0em}{1.7em}\nonumber\\
\al\al J^{\mu\nu}=
(q^\mu q^\nu-g^{\mu\nu} q^2)\,J^{(1)}(t)+q^\mu 
q^\nu\,J^{(2)}(t) \rule{0em}{1.7em}\nonumber\\
\al\al J^{(1)}(t)=\frac{1}{4\, t\,(d-1)}\{(t-4M^2)\,J(t)+2\Dpi\}
\nonumber\\
\al\al J^{(2)}(t)=\frac{1}{4}\,J(t)-\frac{1}{2\,t}\,\Dpi\nonumber\eea
{\bf 1 meson, 1 nucleon:\; $I=I_{11}$}
\bea\al\al \{I,\,I^\mu\}=
\frac{1}{i}\int_{I}\frac{d^dk}{(2\pi)^d}\,\frac{\{1\,,\,k^\mu\}}
{(M^2-k^2)(m^2-(P-k)^2)}\nonumber\\
\al\al s=P^2\co\hspace{2em}\Omega=\frac{s-m^2-M^2}{2M m}\nonumber\\
\al\al I(s)=\bar{I}(s)-\frac{s-m^2+M^2}{s}\,
\bar{\lambda}\co\nonumber\\
\al\al \JSbar(s)=-\frac{1}{8\pi^2}\,\frac{\alpha
\sqrt{\rule{0em}{0.85em}1-\Omega^2}}
{1+2\, \alpha\, \Omega +
  \alpha^2}\; \mbox{arccos}\left(-\frac{\Omega+\alpha}{\sqrt{1+2\, \alpha\, 
\Omega +  \alpha^2}}\right)\nonumber\\
\al\al\hspace{2em} -\frac{1}{16\pi^2}\,\frac{\alpha(\Omega+\alpha)}
{1+2\, \alpha\, \Omega + \alpha^2}
\left(2\,\mbox{ln}\,\alpha-1\right)\co\nonumber\\
\al\al I^\mu=P^\mu I^{(1)}(s)\rule{0em}{1.7em}\nonumber\\
\al\al I^{(1)}(s)=\frac{1}{2s}\,\left\{(s-m^2 + M^2)\,\JS(s) +\Dpi\right\}
\nonumber\eea
{\bf 2 mesons, 1 nucleon:}
\bea \al\al\{I_{21}\,,\,I_{21}^{\;\;\mu}\}=
\frac{1}{i}\!\int_I\frac{d^d k}{(2\pi)^d}\frac{\{1\,,\,k^\mu\}}
{(\Mp^2-k^2)\,
(\Mp^2-(k-q)^2)\,(\Mbare^2-(P-k)^2)}\nonumber\eea
For the low energy analysis of the scalar form factor and of the 
$\pi N$ scattering amplitude to one loop, these integrals are relevant
only on the mass shell of the two external nucleons,
\bea
P^2=P^{\prime\, 2}=
m^2\co\hspace{1em} P'\equiv P-q\co\hspace{2em} 
t=q^2\fs\nonumber\eea
The decomposition of the vectorial integral then
simplifies to
\bea
\al\al I_{21}^{\mu}=(P^\mu+P^{\prime\,\mu})\,I_{21}^{(1)}(t)+
\mbox{$\frac{1}{2}$}
\,q^\mu I_{21}(t)\co\nonumber\\
\al\al I_{21}^{(1)}(t)=\frac{1}{2(4m^2-t)}\left\{(2 M^2-
 t )\,I_{21}(t)-2\,I(m^2)+2\,J(t)\right\}\co\nonumber\eea
and the renormalization of the scalar integral reads
\bea I_{21}(t)=\bar{I}_{21}(t)+\frac{4\,\bar{\lambda}}{\sqrt{t\,(4m^2-t)}}\, 
\arcsin \frac{\sqrt{t}}{2m}\fs\nonumber
\eea
The representation for $\gamma(t)$ constructed in appendix \ref{A} --
which also holds near threshold, but neglects terms beyond next-to-leading
order -- implies
\bea 
\al\al \bar{I}_{21}(t)= 
\frac{1}{32\pi \Mbare \Mp}\left\{\frac{1}{\sqrt{\tau}}\;
\mbox{ln}\,\frac{2 +\sqrt{\tau}}{2 -\sqrt{\tau}}
-\mbox{ln}\left(1+\frac{\alpha}
{\sqrt{4-\tau}}\right)\right\}\nonumber\\\al\al+\frac{1}{32\pi^2 m^2}\left\{
 \frac{2\, (2 -\tau)}{\sqrt{\tau (4 -\tau)}}\;\arcsin
\frac{\sqrt{\tau}}{2}+
\frac{\pi}
{\sqrt{4 -\tau}}+
2\,\mbox{ln}\alpha+1\right\}+O(p)\co\nonumber\eea
with $\tau=t/M^2$.

\vspace{0.5em}\noindent
{\bf 1 meson, 2 nucleons:}
\bea \al\al I_{12}=\frac{1}{i}\!\int_I\frac{d^d k}
{(2\pi)^d}\frac{1}
{(\Mp^2-k^2)\,
(m^2-(P-k)^2)\,(m^2-(P'-k)^2)}\nonumber\eea
In the context of $\pi N$ scattering, 
there are graphs that involve the values of
$I_{12}=I_{12}\,(P^2,P^{\prime\,2},t)$ off the mass shell and
we therefore leave the variables $P^2$, $P^{\prime\, 2}$ open.
The representation (\ref{phimnS}) yields
\bea \al\al I_{12}\,(P^2,P^{\prime\,2},t)
=(4 \pi)^{-\frac{d}{2}}\, \Gamma(3-\mbox{$\frac{d}{2}$})
\int_0^1\!\! dy\!\int_0^\infty \!\! dz\, z\, C^{\frac{d}{2}-3}\co\nonumber\\
\al\al C=M^2(1-z)^2 + z^2\{m^2-y\, (1-y)\, t\}- 2 m M z\,(1-z)\,
\bar{\Omega}\co\nonumber
\\ \al\al t=(P-P')^2\co\hspace{5em} \bar{\Omega}=
y\,\Omega +(1-y)\,\Omega'\co\nonumber\\
\al\al \Omega=\frac{P^2-m^2-M^2}{2\,m M}\co\hspace{2em} \Omega'=
\frac{P^{\prime\,2}-m^2-M^2}{2\,m M}\fs\nonumber\eea
The expression for $C$ shows that the momentum transfer enters exclusively 
through $m^2-y\,(1-y)\,t$.
The expansion in powers of $t$ thus only yields
polynomials -- the phenomenon observed in the case of $\gamma(t)$ does not
occur here. Removing the pole term at $d=4$ (see section
\ref{renormalization}.1), scaling the variable of integration with
$z\rightarrow \alpha \, u$ and expanding in powers of $\alpha$, we
obtain
\bea
\al\al I_{12}\,(P^2,P^{\prime\,2},t)=
\bar{I}_{12}\,(P^2,P^{\prime\,2},t)+\nu_{12}\,\bar{\lambda}
\co\nonumber\\
\al\al\bar{I}_{12}\,(P^2,P^{\prime\,2},t)=-\frac{1}{16\pi^2 m^2}
\left\{\frac{f_1(\Omega)-f_1(\Omega')}
{\Omega-\Omega'}\right\}+\frac{\nu_{12}}{32\pi^2}\,
\left\{2\,\mbox{ln}\,\alpha+1\right\}
+O(p^2)\nonumber\\
\al\al\nu_{12}=-\frac{1}{m^2}\left\{1-\alpha\,\Omega-\alpha\,\Omega'+
O(p^2)\right\}\nonumber\\
\al\al f_1(\Omega)=(1-\alpha\,\Omega)\,\sqrt{1-\Omega^2}\;\;
\mbox{arccos}\,(-\Omega)-\Omega\nonumber\fs\eea
At leading and first nonleading order, the result is independent of the 
momentum transfer. On the mass shell, it reduces to 
\bea \bar{I}_{12}(t)\equiv\bar{I}_{12}\,(m^2,m^2,t)=-\frac{1}{32\pi^2 m^2}\,
(2\,\mbox{ln}\,\alpha+1)+\frac{\alpha}{64\pi m^2}+O(p^2)\fs\nonumber \eea
{\bf 1 meson, 3 nucleons:} 
\bea I_{13}\!=\!\frac{1}{i}\!\int_I\!\frac{d^d k}
{(2\pi)^d}\frac{1}{(\Mp^2-k^2)
(m^2-(P_1-k)^2)(m^2-(P_2-k)^2)(m^2-(P_3-k)^2)}\nonumber\eea
The integral is relevant for $\pi N$ scattering. Denoting the momenta of the
two incoming and outgoing particles by $P,q$ and $P',q'$, respectively, we
have
\bea\al\al 
P_1=P\co\hspace{1em}P_2=P+q\co\hspace{1em}P_3=P'-q'\co\hspace{1em} 
P^2=P^{\prime\,2}=m^2\co\nonumber\\\al\al q^2=q^{\prime\,2}=M^2
\co\hspace{1em}
s=(P+q)^2=m^2 +
  M^2 +2 M m\, \Omega\co\hspace{1em}t=(q-q')^2\fs\nonumber\eea
The chiral expansion starts at order $1/p$. The coefficients are readily worked
out with the representation (\ref{phimnS}), setting $d=4$. As in the case of
$I_{12}$, the leading and first nonleading terms are independent of $t$:
\bea\al\al I_{13}=-\frac{1}{32\pi^2 M
  m^3\,\Omega^2}\,f_0(\Omega)+\frac{1}{32\pi^2 
  m^4\,\Omega^3}\left\{f_0(\Omega)+\frac{\pi}{4}\,\Omega^2+\Omega^3
\right\}+O(p)\nonumber\\
\al\al f_0(\Omega)=\sqrt{1-\Omega^2}\,\arccos(-\Omega)-\frac{\pi}{2}-\Omega
\nonumber\fs\eea

\section{$\pi N$ scattering in tree approximation}\label{scattering}
\setcounter{equation}{0}
In connection with the unitarity condition for the scalar form factor, 
we need the representation of the $\pi N$ scattering amplitude to order $p^2$.
For this purpose, the standard decomposition into the invariant 
amplitudes $A$ and $B$,
is not suitable, because the leading 
contributions from the 
two cancel at low 
energies\footnote{An evaluation of the scattering
amplitude to $O(p^n)$ suffices to determine $D(\nu,t)$ to the same
order and yields $B(\nu,t)$ to $O(p^{n-2})$, 
but fixes $A(\nu,t)$ only to $O(p^{n-1})$. Conversely, a representation for
the pair $(D,B)$ to accuracy $(p^n,p^{n-2})$ fully characterizes the
scattering amplitude to $O(p^n)$, while an analogous representation for the
pair $(A,B)$ does not suffice, even if $A$ is assumed known to all orders.}. 
We replace $A$ by 
$D\equiv A+\nu\,B$,
\bea T=\bar{u}' \left\{A +\frac{1}{2}\,(q\hspace{-0.5em}/^{\,\prime}+
q\hspace{-0.5em}/\hspace{0.1em} )B\right\} u=
\bar{u}'\left\{ D -\frac{1}{4m} [\,q\hspace{-0.5em}/^{\,\prime},\,
q\hspace{-0.5em}/\hspace{0.15em}]B \right\}u \co 
\nonumber\eea
and work with the variables
\bea \nu\al=\al\frac{s-u}{4\,
  m}\co\hspace{3em}\nu_B=\frac{t-2 M^2}{4\, m}\fs\nonumber\eea 
The tree graphs of the Lagrangian ${\cal
  L}_N^{(1)}+{\cal L}_N^{(2)}$ yield
\bea\label{C1} 
D^+(\nu,t)\al=\al\frac{g_A^2\,m}{F^2}\;\frac{\nu_B^2}{\nu_B^2-\nu^2}-
\frac{4\,c_1 M^2}{F^2}+\frac{2\,c_2\,\nu^2}{ F^2}+
\frac{c_3\, (2M^2-t)}{F^2}+O(p^3)\co\nonumber\\
B^+(\nu,t)\al=\al
\frac{g_A^2\,m}{F^2}\;\frac{\nu}{\nu_B^2-\nu^2}+O(p)\co\\
D^-(\nu,t)\al=\al \frac{g_A^2\, m}{F^2}\,\frac{\nu\;\nu_B}{\nu_B^2-\nu^2}-
\frac{\nu}{2F^2}\,(g_A^2-1)+O(p^3)\co\nonumber\\
B^-(\nu,t)\al=\al\frac{g_A^2\, m}{F^2}\,\frac{\nu_B}{\nu_B^2-\nu^2}-
\frac{1}{2F^2}\,(g_A^2-1)+\frac{2\,c_4\,m}{F^2}+O(p)\fs\nonumber \eea
The projection onto the $t$-channel $I=J=0$ partial wave
reads
\bea \al\al\hspace{-1em} f^0_+(t)= f_{\!B+}^{\,0}(t)+\frac{m^2}{24\pi F^2}
\!\left\{-24M^2 c_1+(4M^2-t)\,c_2 +
6\, (2M^2-t)\,c_3 \right\}\!+O(p^3)\,,\nonumber\\
\al\al\hspace{-1em} f_{\!B+}^{\,0}(t)= \frac{g_A^2\, m^3}
{4\pi F^2}\left\{
\frac{\arctan\,\tau}{\tau}-\frac{t}{4m^2}\right\}
\co\label{fBorn}\hspace{1.5em}\tau= 
\frac{\sqrt{(t-4M^2)(4m^2-t)}}{t-2M^2}\fs\eea
The comparison with the subthreshold expansion of
H\"ohler et al. \cite{hoe},
\bea D^+(\nu,t)\al=\al\frac{g_{\pi N}^2}{m}\;
\frac{\nu_B^2}{\nu_B^2-\nu^2}+d_{00}^+
+d_{10}^+\,\nu^2 +d_{01}^+\,t+d_{20}^+\,\nu^4+\ldots \nonumber\\
B^-(\nu,t)\al=\al\frac{g_{\pi N}^2}{m}\;
\frac{\nu_B}{\nu_B^2-\nu^2}-\frac{g_{\pi N}^2}{2\,m^2}+b_{00}^-
+b_{10}^-\,\nu^2 +b_{01}^-\,t+\ldots\nonumber \eea
implies the following representations\footnote{For a
  discussion of the  
corrections of order $M$, we refer to section \ref{formfactor}.5.} for the 
coupling constants \cite{BKM NPB457}:
\bea \label{ctree} c_1\al=\al -\frac{F^2}{4 M^2}\,
 \left\{d_{00}^+ +2 M^2\,d_{01}^+
\right\}+O(M)\co\hspace{1em}
c_2= \frac{F^2}{2}\; d_{10}^++O(M)\\
c_3\al=\al - F^2\, d_{01}^++O(M)\co\hspace{8em}
c_4=\frac{1}{2 m}\left\{F^2\,b_{00}^- -\mbox{$\frac{1}{2}$}\right\} 
+O(M)\fs\nonumber\eea
The numerical values $d_{00}^+=-1.46$, $d_{10}^+=1.12$, $d_{01}^+=1.14$,
$b_{00}^-=10.36$ (in pion mass units, \cite{hoe}) thus lead 
to \bea c_1^{(0)}= -0.60\,m_N^{-1}\co\hspace{1em}c_2^{(0)}= 1.6\,m_N^{-1}\co
\hspace{1em}
c_3^{(0)}= -3.4\,m_N^{-1}\co\hspace{1em} c_4^{(0)}= 2.0\,m_N^{-1}\fs
\nonumber\eea
The current algebra formula $\sigmaN=-4\, c_1 M^2$ then yields $\sigmaN=50
\,\mbox{MeV}$. Alternatively, replacing the input for $d_{00}^+$ by  
$\sigmaN=45\,\mbox{MeV}$ \cite{GLS}, we obtain $c_1^{(0)}
= - 0.54\,m_N^{-1}$ and $d_{00}^+=-1.54$.

\setcounter{equation}{0}
\section{Contributions generated by the $\Delta$}\label{delta}

\subsubsection*{Effective Lagrangian}\label{delta1}
The $\Delta$ degrees of freedom may be described in terms of
a Rarita-Schwinger spinor $\psi^a_\mu$, which transforms with the 
representation $D^{\frac{1}{2}}\times D^1$ of the isospin group. The condition 
$\tau^a\psi^a_\mu=0$ eliminates the component with 
$I=\frac{1}{2}$.
To lowest order, the relevant effective Lagrangian reads
\begin{eqnarray}\label{LDelta}
  {\cal L}_\Delta\al =\al\bar\psi^a_\mu
  \Lambda^{\mu\nu}\psi^a_\nu+ 
g_{\Delta} F\,
(\bar\psi^a_\mu\,
u^a_\nu\,  \Theta^{\mu\nu}\, \psi+ \mbox{h.c.})
\co\eea
with $u^a_\mu=\frac{1}{2}\,\mbox{tr}\,(\tau^a u_\mu)$.
The kinetic term involves the tensor
\bea \Lambda^{\mu\nu}=-(iD\hspace{-0.7em}/-m_\Delta)g^{\mu\nu}
+i(\gamma^\mu D^\nu+\gamma^\nu D^\mu)-
\gamma^\mu(iD\hspace{-0.7em}/+m_\Delta)\gamma^\nu \nonumber\co\eea
while the one occurring in the interaction is of the form
\bea \Theta^{\mu\nu}=g_{\mu\nu}-(\mbox{$\frac{1}{2}$}+Z)\gamma_\mu \gamma_\nu 
\fs\nonumber
\end{eqnarray}
We use the notation of ref.\cite{hoe}, where the coupling constant $g_\Delta$
carries the dimension of an inverse mass\footnote{In the
  literature, the coupling constant $g_\Delta$ is often replaced
by the dimensionless quantity $F_\pi g_\Delta = C/\sqrt{2} \;
\cite{BKMSigma,Manohar}  = h_A \;\cite{Tang}
= g_{\pi N\Delta}\;\cite{Hemmert} $.}. 
The expression for the width of the $\Delta$ 
that follows from the tree graphs
of the above Lagrangian reads  
\bea \Gamma_{\Delta\rightarrow N\pi}=\frac{g_\Delta^2
  q_\Delta^3}{24 \pi\, m_\Delta^2}\,\left\{(m_\Delta+m)^2-M^2\right\}\co
\nonumber\eea 
where $q_\Delta$ is the momentum of the decay products for the decay at rest.
In the numerical evaluation, we use $m_\Delta=1.232\, \text{GeV}$ and
the value of the coupling constant advocated in ref.~\cite{hoe}  
$g_\Delta=13.0\,\text{GeV}^{-1}$, 
$F_\pi\, g_\Delta = 1.2$. There is considerable scattering in the numerical
values used in the literature. In particular, the experimental
width implies 
stronger coupling -- the significance of the difference is discussed in
detail in ref.~\cite{hoe}, on the basis of an analysis of the 
$P_{33}$-wave. Some of the values found within the 
small scale expansion instead indicate a weaker coupling 
(see ref.~\cite{Hemmert} for a review of that expansion 
in HB$\ch$PT and ref.~\cite{Tang}
for the corresponding relativistic formulation).

\subsubsection*{Contribution to the $\pi N$ scattering amplitude}\label{delta2}
In addition to the coupling constant $g_\Delta$ that determines the residue of
the poles arising from $\Delta$-exchange, the above effective
Lagrangian contains the parameter $Z$, which generates 
polynomial contributions to the scattering amplitude
(the representation obtained from the tree graphs of this 
Lagrangian can be found in ref.~\cite{hoe}). To fix the
value of $Z$, we observe that the unitarity bounds \cite{Sommer} ensure
an unsubtracted dispersion relation for $B^+(\nu,0)/\nu$. The relation
represents this amplitude
as a superposition of contributions arising from the various intermediate
states. While the nucleon generates the
Born term in eq.~(\ref{C1}), the $\Delta$ produces two peaks in the 
imaginary part, at $\nu=\pm\,\omega_\Delta$, where
$\omega_\Delta=(m_\Delta^2-m^2-M^2)/2 m$. In the narrow width approximation,
the corresponding contribution to the dispersion integral is of the form
\bea \frac{1}{\nu}\,B^+_\Delta(\nu,0)= b_0
\left\{\frac{1}{\omega_\Delta-\nu}+\frac{1}{\omega_\Delta+\nu}\right\}
\nonumber\fs\eea
The tree graph amplitude generated by the above effective Lagrangian 
does contain such pole terms, but involves an
additive constant proportional to $Z^2$. In the language of the dispersion
relation, this term corresponds to the integral over those
intermediate states that remain when the nucleon and the $\Delta$ are removed.
In order for the effective Lagrangian in
eq.~(\ref{LDelta}) to properly account for the
contributions generated by the $\Delta$, we must set $Z=0$.

The corresponding expressions for the invariant amplitudes read \cite{hoe}:
\begin{eqnarray*}
    A^{\pm}_{\Delta}(\nu,t)\al=\al \frac{{\displaystyle
 g_{\Delta}^2}}{{\displaystyle 18 m}}\, 
 \left\{\dbinom{2}{-1}\,(\alpha_1+\alpha_2\, t) \,
  \left(\frac{1}{\nu_\Delta-\nu}\pm
         \frac{1}{\nu_\Delta+\nu}\right)+\dbinom{\alpha_3}{0}\right\} \co\\ 
   B^{\pm}_{\Delta}(\nu,t)\al=\al \frac{g_{\Delta}^2}{18 m}
\left\{\binom{2}{-1}\,(\beta_1+\beta_2\, t)\, 
   \left(\frac{1}{\nu_\Delta-\nu}\mp
 \frac{1}{\nu_\Delta+\nu}\right)+\binom{0}{\beta_3}\right\} \co
\end{eqnarray*}
with $\nu_\Delta=\omega_\Delta+t/4 m$. The coefficients are fixed by the 
masses. The expressions may be simplified by using the parameter 
$E_\Delta=( m_\Delta^2+m^2-M^2)/2 m_\Delta$ (energy of the nucleon 
in the rest frame of the $\Delta$):
\begin{align*}
  \alpha_1 &
  =2\,(E_\Delta+m)\,\{E_\Delta\,(2\,m_\Delta+m)-m\,(m_\Delta+2\,m)\}
  \co\nonumber\\ 
 \alpha_2 & =3\,(m_\Delta+m)/2\co \\
 \alpha_3 & =-4\, m\,(m_\Delta+m)\,(2\, m_\Delta^2+m\, 
           m_\Delta-m^2+2\,M^2)/m_\Delta^2\co\\  
  \beta_1&=2\,(E_\Delta+m)(E_\Delta-2\,m)\co \\
 \beta_2&= 3/2\co\\
  \beta_3&=2\,m (m_\Delta+m)^2/m_\Delta^2\fs 
\end{align*}
For the reason given in appendix \ref{scattering},
the amplitudes $A^\pm$ are not suitable to sort out the low energy structure. 
The two terms in 
$D^\pm=A^\pm +\nu\,B^\pm$ nearly cancel also in the present case:
The chiral expansion of $D^+_{\!\Delta}$ and $D^-_{\!\Delta}$ only starts 
at $O(p^2)$ and $O(p^3)$, respectively. We do not list the corresponding
explicit representations -- these are readily obtained from the above formulae.

\subsubsection*{Subtraction constants and saturation}\label{delta3}
The QCD Lagrangian does not contain parameters that explicitly refer to the
$\Delta$, nor is this state singled out by chiral symmetry. 
We are identifying the $\Delta$-contributions on the basis of 
a saturation hypothesis: 
The argument used to pin down these contributions relies on the fact 
that the dispersion relation
for $B^+(\nu,t)/\nu$ does not contain a subtraction -- we have determined
the contributions from the $\Delta$ by saturating this relation for $t=0$. 
We now briefly discuss the dispersion relations obeyed by the three 
other amplitudes from the same point of view. More specifically, 
we wish to show that
\begin{itemize}
\item The dispersion relation for $D^+(\nu,t)$ contains a subtraction, 
which ensures that this amplitude has an Adler zero in the vicinity
of the point $\nu=0$, $t=2M^2$.
\item The amplitude $D^-(\nu,t)/\nu$ obeys an unsubtracted dispersion 
relation, which is closely related to the
Adler-Weisberger relation and allows us to identify the $\Delta$-contribution
to the coupling constant $g_A$.
\item Finally, $A^-(\nu,t)/\nu$ also obeys an unsubtracted dispersion
relation. The representation for this amplitude obtained by saturating the 
dispersion integral with the $\Delta$ is identical with the one given above.
This confirms that the parameter $Z$ must be set equal to zero.
\end{itemize}
For simplicity, we only consider the forward dispersion relations, i.e.~set 
$t=0$ (in the general case, the subtraction ``constants'' are functions of
$t$). As the unitarity bounds of ref.~\cite{Sommer} 
are too weak to discuss the subtractions occurring
in the dispersion relations for the isospin odd amplitudes,
we invoke Regge pole analysis, which 
yields more specific information about the asymptotic behaviour. 

In the case of the isospin even amplitudes, 
the asymptotics is dominated by Pomeron exchange: $A^+(\nu,t),
D^+(\nu,t) \propto
\nu^{\alpha_P(t)}$, $B^+(\nu,t)\propto \nu^{\alpha_P(t)-1}$, with
$\alpha_P(0)=1$ (we disregard the logarithmic terms arising from the fact
that the Pomeron does not represent a simple Regge
pole). Since the Regge behaviour is consistent with the unitarity bounds,
it merely confirms that $B^+(\nu,0)/\nu$
obeys an unsubtracted dispersion relation. The
dispersive representation for $D^+(\nu,0)$, however, requires a 
subtraction (note that $D^+(\nu,t)$ is even in $\nu$, so that a single
subtraction suffices). In fact, if this amplitude were to obey an unsubtracted
dispersion relation, the saturation with the $\Delta$ would lead to a
conflict with chiral symmetry, which implies that
$D^+$ contains an Adler zero. In the above explicit representation, 
the zero arises because the 
subtraction constant $\alpha_3$ compensates the pole terms: 
$D^+_{\!\Delta}(0,2 M^2)$ represents a term of order $M^4$. The
occurrence of a subtraction is essential for this to happen.

For the isospin odd amplitudes, the $\rho$-trajectory
yields the leading term: $A^-(\nu,t),D^-(\nu,t)\propto \nu^{\alpha_\rho(t)}$,
$B^-(\nu,t)\propto \nu^{\alpha_\rho(t)-1}$, with $\alpha_\rho(0)\simeq
\frac{1}{2}$. This implies that the
functions 
\bea \hat{A}(\nu)=\frac{1}{\nu}\,A^-(\nu,0)\co\hspace{2em}
\hat{D}(\nu)=\frac{1}{\nu}\,D^-(\nu,0)\nonumber\eea
obey unsubtracted dispersion relations 
(the same also applies for $B^-(\nu,0)$,
which represents the difference between the two). 
 Indeed, the representation given for  
$A^-_\Delta(\nu,0)$ is precisely the one obtained by 
saturating the unsubtracted dispersion relation for $\hat{A}(\nu)$ with 
the $\Delta$. The one for $D^-_{\!\Delta}(\nu,0)$, however, does contain a
subtraction term, $\beta_3$, in apparent conflict with the statement that
this amplitude obeys an unsubtracted dispersion relation. 

\subsubsection*{Adler-Weisberger relation}\label{Adler-Weisberger}
To resolve the paradox, we consider the dispersion relation for
$\hat{D}(\nu)$, which according to the above is free of subtractions:
\bea \hat{D}(\nu)=\frac{g_A^2 m}{F^2}\,\frac{ \nu_B}{\nu_B^2-\nu^2}+
\frac{2}{\pi}\int_M^\infty
\frac{d\nu'}{\nu^{\prime \,2}-\nu^2}\;\mbox{Im}\,D^-(\nu',0)\fs\nonumber
\eea 
We have explicitly indicated the contribution from the one nucleon
intermediate state, with $\nu_B=-M^2/2\,m$.
The comparison with the low energy theorem (\ref{C1})
leads to an on-shell variant of the Adler-Weisberger 
relation:
\bea \frac{1}{2F^2}\,(g_A^2-1)+\frac{2}{\pi}\int_M^\infty
\frac{d\nu'}{\nu^{\prime \,2}}\;\mbox{Im}\,D^-(\nu',0)=O(p^2)\fs\nonumber
\eea
The imaginary part may be expressed in terms of total cross sections.
Dropping the corrections\footnote{For an analysis
of the terms of $O(p^2)$, we refer to Brown, Pardee and Peccei 
\cite{Brown Pardee Peccei}.} of order $p^2$ and replacing $F$ by $F_\pi$, the
result obtained for the dispersion integral yields a
measurement of $g_A$, based on total $\pi N$ cross sections. 
The result is in remarkably good agreement with the
value found from neutron decay \cite{hoe}. 

Replacing the dispersion integral by the contribution from the
$\Delta$, the above formula yields
\bea g_A^{2}=1+
\frac{2}{9}\,F^2 g_\Delta^2\,\frac{\{(m_\Delta+m)^2-M^2\}^2
\,\{(m_\Delta -m)^2 -M^2\}}{ m_\Delta^2\,(m_\Delta^2-m^2-M^2)^2}\fs
\nonumber\eea
The corresponding numerical value, $g_A=1.35$, shows that the contribution
from the $\Delta$ indeed dominates the dispersion
integral: The remainder is about four times smaller (and of opposite sign). 

This resolves the paradox encountered above: The explicit representations for
the amplitudes $D_\Delta^\pm$, $B_\Delta^\pm$ do not
include the effect generated by the $\Delta$ in the value of the coupling
constant $g_A$, which is treated as an independent parameter -- 
this is why the representations for $D^-_{\!\Delta}$ and $B_{\Delta}^-$ 
involve the subtraction term $\beta_3$, despite the fact that
the full amplitudes obey unsubtracted dispersion relations. The difference
only concerns these two amplitudes: The formulae (\ref{C1}) show that
the polynomial term proportional to $g_A^2-1$, which causes the
paradox, only enters the expressions 
for $D^-$ and $B^-$ (in $A^-=D^--\nu\,B^-$, the term drops out).

\subsubsection*{Subthreshold expansion}\label{delta4}

The $\Delta$-contributions to the coefficients of the 
subthreshold expansion are obtained by
expanding the amplitudes in powers of 
$\nu$ and $t$, with the result
\begin{align}
d_{00\Delta}^+&=\frac{g_{\Delta}^2}{18\, m\,\omega_\Delta}\,(4
  \alpha_1 +\omega_\Delta  \alpha_3)\co & 
d_{10\Delta}^+&=
 \frac{2\, g_{\Delta}^2 }{9
 \, m\, \omega_\Delta^3}\,(\alpha_1+\omega_\Delta \beta_1)\co \nonumber\\
d_{01\Delta}^+&=
- \frac{g_{\Delta}}{18\,
  m^2\, \omega_\Delta^2}\,(\alpha_1-4 \alpha_2 m \omega_\Delta)\co &
b_{00\Delta}^-&=
-\frac{g_{\Delta}^2}{18\, m\,\omega_\Delta}\,
(2 \beta_1-\omega_\Delta \beta_3)\fs\nonumber
\end{align}
Numerically, this amounts to $(d_{00\Delta}^+,d_{10\Delta}^+,
d_{01\Delta}^+,b_{00\Delta}^-)=(-1.4,\,0.8,\,0.7,\,5.3)$, in units of
$M_\pi$. The 
comparison with the phenomenological values of ref.~\cite{hoe},  
$(d_{00}^+,d_{10}^+,
d_{01}^+,b_{00}^-)=(-1.46,\, 1.12,\, 1.14,\,10.36)$, 
shows that the $\Delta$ indeed
accounts for a significant part of the curvature seen
in the scattering amplitude at small values of $\nu$ and $t$. The
situation is similar to the one in the mesonic sector, where the most
important singularity that is not explicitly accounted for in the effective
theory is the one generated by the $\rho$.

It is instructive to compare the chiral expansion of the subthreshold 
coefficients with the small scale expansion thereof. 
In the chiral expansion, 
the coefficients are expanded in powers of $M$ at fixed $m_\Delta$. The small
scale expansion results if the mass difference $\Delta=m_\Delta-m$ is also 
treated as a small quantity and the double series in $M$ and $\Delta$
is reordered by counting the ratio $M/\Delta$ 
as a quantity of order 1. The chiral
expansion yields a decent approximation for all of the above coefficients
(the largest deviation occurs in $d_{10}^+$, but it still amounts to less than
20 \%). The small scale expansion does not improve the accuracy of this
approximation -- in fact, the discrepancy between the leading term of that
expansion and the full result for $d_{10}^+$
is about twice as large as for the chiral series. 
The numerical exercise shows
that -- as long as we stick to small values of $\nu$ and $t$ -- the
singularities generated by the $\Delta$ do not jeopardize the chiral
perturbation series. 

\subsubsection*{Contribution to the effective coupling constants of 
${\cal L}_N^{(2)}$} \label{delta5}

To read off the contributions of the $\Delta$ to the effective coupling
constants of ${\cal L}^{(2)}_N$, we expand the above expressions 
in powers of $M$ and insert the result in eqs.~(\ref{cd}). 
This leads to\footnote{Note that the coupling constant $c_4$ picks
up a small negative extra contribution from the term 
$c^{\scriptscriptstyle F}_4=-1/4m$ in
eq.~(\ref{ctree}).}:
\bea
\al\al c_1^\Delta=0 \co \hspace{1.5em}
c_2^\Delta= 2\,c\, m\,(m_\Delta+m)\co \hspace{1.5em}
c_3^\Delta=-c\,(4m_\Delta^2-m_\Delta m+m^2)\co \nonumber\\ 
\al\al c_4^\Delta=c\,(2\,m_\Delta-m)(m_\Delta +m)\co 
\hspace{3em} c\equiv\frac{g_\Delta^2
  F^2}{9\,m_\Delta^2(m_\Delta-m)}\fs\label{couplings} \eea
Numerically, these formulae yield
$(c_1^\Delta,c_2^\Delta,c_3^\Delta,c_4^\Delta)=
(0,\,1.4,\,-2.0,\,1.1)\,m_N^{-1}$.
Again, the higher order terms of the chiral expansion 
do not significantly modify these estimates: Evaluating the relations
(\ref{cd}) with the full 
expressions for the coefficients of the subthreshold
expansion instead of only the leading terms, we obtain 
$(c_1^\Delta,c_2^\Delta,c_3^\Delta,c_4^\Delta)=
(-0.04,\,1.2,\,-2.1,\,1.2)\,m_N^{-1}$. In particular, 
the higher order effects generate a small contribution to the coupling
constant $c_1$, which amounts to a shift in the 
$\sigma$ term by about 3 MeV.

\subsubsection*{Contribution to $\Delta_\sigma$ and to the coupling constants 
$\bar{e}_1$, $\bar{e}_2$}\label{delta6}
As discussed in section \ref{formfactor}.4, the corrections to the leading
order result for the difference $\Delta_\sigma=\sigma(2 M^2)-\sigma(0)$
are dominated by the logarithmic term, $\Delta_2\, M^4\ln M/m$. 
A significant fraction of this term is
generated by the $\Delta$: Inserting the above expressions in 
eq.~(\ref{Deltasigmac}), we obtain
\bea\label{Delta2Delta} \Delta_2\;\rule[-0.5em]{0.03em}{1.3em}
\,\rule[-0.4em]{0em}{0em}_\Delta =-
\frac{g_\Delta^2 (6\,m_\Delta^2
-2\,m_\Delta m +m^2) }{36 \pi^2\,m_\Delta^2(m_\Delta-m)}\fs\eea
Numerically, this amounts to $\Delta_2\, M^4\ln M/m=\mbox{5.9 MeV}$, to be 
compared with the
number 10.6 MeV, obtained by inserting the phenomenological values for the 
combination $c_2 + 6\,c_3$ of effective coupling constants (the contribution
from $g_A$ is of opposite sign and reduces this to 7.8 MeV). 

The effects generated by the $\Delta$ in the scalar form factor were analyzed
quite some time ago \cite{BKMSigma}. That calculation is equivalent to an
evaluation of the dispersion integral (\ref{dIDelta}) in the approximation
where the integrand is replaced by the leading term of the small scale
expansion. The corresponding representation for the contribution to the
partial wave $f^0_+(t)$ generated by $\Delta$ exchange reads \bea
f^{\,0}_{\!\Delta+}(t)=\frac{2\,g_\Delta^2 m^2}{9\pi}\left\{ \frac{2\,\Delta^2
    + t - 2\,M^2}{\sqrt{t-4\,M^2}}\,\arctan\,
  \frac{\sqrt{t-4\,M^2}}{2\,\Delta}-\Delta\right\}\co\nonumber\eea with
$\Delta\equiv m_\Delta-m$. At low energies, this approximation yields a
remarkably good representation for the full partial wave projection of the
$\Delta$-contribution to the scattering amplitude.  The leading term in the
chiral expansion of the dispersion integral then reads \cite{Kambor}:
\begin{equation}
\Delta_\sigma\;\rule[-0.5em]{0.03em}{1.3em}
\,\rule[-0.4em]{0em}{0em}_\Delta=\frac{g_\Delta^2 M^4}{36\pi^2\Delta}
\left\{\frac{5}{3}-\frac{\pi}{4}-
5 \ln\left(\frac{M}{2\,\Delta}\right)+
O\left(\frac{M^2}{\Delta^2}\right)\right\}
\fs\nonumber
\end{equation}
The coefficient of the chiral logarithm occurring here indeed agrees with 
the leading term in the small scale expansion of the representation
(\ref{Delta2Delta}). Note that there is a difference between the 
chiral perturbation series and the small scale expansion: In the former,
chiral logarithms only start showing up at $O(p^4)$, but for the latter,
those generated by the $\Delta$ manifest themselves already at third order.

The remainder of the above representation amounts to an estimate for
the $\Delta$-contribution to the coupling constant $e_2$:
\bea \bar{e}_2^\Delta
=\frac{g_\Delta^2}{144\pi^2\, \Delta}\left\{10\, \ln
  \left(\frac{2\,\Delta}{m}\right)+7+O\left(\frac{M^2}{\Delta^2}\right)\right\}
\fs\nonumber\eea
The formula confirms that the coupling constant $\bar{e}_2$ is small:
Numerically, the corresponding contribution to $\Delta_\sigma$ amounts
to 0.7 MeV.

The analogous estimate for the $\Delta$-contribution to the low energy
constant $e_1$ may be obtained from the small scale expansion of 
the nucleon mass \cite{Kambor}. The comparison 
with eq.~(\ref{mN}) yields
\bea \bar{e}_1^\Delta=-\frac{g_\Delta^2}{48\pi^2 \Delta}\left\{6\,\ln\left(
\frac{\Delta}{2\,m}\right)+5+O\left(\frac{M^2}{\Delta^2}\right)\right\}
\fs\nonumber\eea 
Numerically, this corresponds to a contribution to the $\sigma$-term of
5.7 MeV. 

Note that, if the degrees of freedom of
the $\Delta$ occur as dynamical variables of the effective theory, 
the quantum fluctuations also generate graphs that 
contain $\Delta$ propagators instead
of nucleon propagators. The small scale expansion allows a coherent counting
of powers also for these graphs. The width of the
$\Delta$, for instance, represents a term of third order, so that
the corresponding shift of the pole in the propagator does
represent a correction which perturbation theory is able to cope with.
There is a difference to the framework described above, insofar as
the constants $m$ and $c_1$ now pick up renormalization.
At leading order, the $\Delta$ is degenerate with the nucleon, so that
the loop graphs containing $\Delta$ propagators also give rise to infrared
singularities. The method described in the present paper may be extended to
analyze these \cite{Tang}.

\end{document}